\documentclass[preprint,aps]{revtex4}

\usepackage{tabularx}
\usepackage{bm}
\usepackage{euscript}
\usepackage{epsfig,psfrag,subfigure}
\usepackage{graphicx}
\usepackage{color}
\usepackage{amsfonts}
\usepackage{exscale}
\usepackage{amsbsy}
\usepackage{stmaryrd}
\usepackage{wasysym}

\def\be{\begin{equation}}       \def\ee{\end{equation}}
\def\bea{\begin{eqnarray}}      \def\eea{\end{eqnarray}}
\def\ba{\begin{array} }
\def\ea{\end{array} }
\def\bnum{\begin{enumerate} }
\def\enum{\end{enumerate}}

\def\=>{\Rightarrow}
\def\>{\rightarrow}

\def\eye2{Fathbb{I}}

\begin{document}

\author{Steven~A.~Kivelson, Andrew~C.~Yuan }
\affiliation{Department of Physics, Stanford University, Stanford, CA 94305, USA,\\
Stanford Institute for Materials and Energy Sciences, SLAC National Accelerator Laboratory, 2575 Sand Hill Road, Menlo Park, CA 94025}
\author{B.~J. Ramshaw}
\affiliation{Laboratory of Atomic and Solid State Physics, Cornell University, Ithaca, NY, 14853}
\author{Ronny Thomale}
\affiliation{Institut f\"ur Theoretische Physik und Astrophysik and W\"urzburg-Dresden Cluster of Excellence ct.qmat, Julius-Maximilians-Universit\"at W\"urzburg, Am Hubland 97074 W\"urzburg, Germany}
\title{A proposal for reconciling diverse experiments on the superconducting state in Sr$_2$RuO$_4$}
\date{\today}

\begin{abstract}
A variety of precise experiments have been carried out to establish the character of the superconducting state in Sr$_2$RuO$_4$.  Many  of these appear to imply contradictory conclusions concerning the symmetries of this state.  Here, we propose that these results can be reconciled if we assume that there is a near-degeneracy between a $d_{x^2-y^2}$ (B$_{1g}$ in group theory nomenclature) and a $g_{xy(x^2-y^2)}$ (A$_{2g}$) superconducting state.
%, and there is sufficient random shear-strain to weakly destabilize a superconducting nematic phase.  
From a weak-coupling perspective, such  an accidental degeneracy can occur at a point at which a balance between the on-site and nearest-neighbor repulsions triggers a $d$-wave to $g$-wave transition.
\end{abstract}

\maketitle

\section{Introduction}
Sr$_2$RuO$_4$ should be the poster child for the theory of ``unconventional'' superconductors: it is a clean, stoichiometric crystalline material, it settles into a Fermi liquid ``normal state'' in a range of temperatures, $T$, that extends roughly a factor of 20 above the superconducting $T_c$,
%\brad{(Just as a point of conversation, I really think the only temperature scale that can be defined in the transport data is one where the c-axis resistivity starts to decrease, around 100 Kelvin. I think claims that this temp scale is \~20 to \~30 K are based on where the resistivity deviates by eye from $T^2$ on a log-scale, which really isn't very meaningful. This is probably more important for something Sayak and I are looking at at the moment - the elastoresistivity (which is indeed large and divergent above about 60 K, then rolls over at lower temp and decreases. I would be interested to know if there is other evidence for a 30 K temperature scale, though}, 
 the normal state fermiology is well established, the superconducting transition is sharp and of mean field type,  and there is no doubt that the superconducting state is unconventional in the sense of a sign-changing order parameter with a mean equal to or near 0 along the Fermi surface~\cite{firstpaper,ishida1997anisotropic,mackenzieunconventional,sroreview}. Of all currently known unconventional superconductors, Sr$_2$RuO$_4$ is the one in which a BCS-like weak coupling mean-field-theory treatment is most easily justified, meaning that the microscopic ``mechanism'' of superconductivity  should be well defined.  Instead, a large number of careful experiments have lead to a set of observations which cannot be accounted for even at the scenario level by any current theory -- %\ronny{
 35 years after the first phenomenological theories~\cite{Rice,Baskaran}
  we are still debating the symmetry of the superconducting order parameter~\cite{sroreview}. 

Indeed, some sets of observations almost appear to be mutually contradictory:  The evidence~\cite{musr,kerr,Kidwingira1267}
%josephson} 
 of time-reversal symmetry breaking (TRSB) in the superconducting (SC) state is difficult to reconcile with the existence of gapless nodal lines in the quasiparticle spectrum~\cite{doi:10.1143/JPSJ.69.572,penetrationdepth,lupien2001ultrasound,thermalconductivity,Irmo,stm2}. The evidence of a two component superconducting order parameter~\cite{lupien2002ultrasound,bradspaper} is difficult to  reconcile with the  lack of singular $T_c$ dependence on  symmetry breaking strain~\cite{hicksstrain} or in-plane magnetic field~\cite{doi:10.1143/JPSJ.71.2839}.
 The evidence of triplet (odd parity~\cite{josephson}) pairing is difficult to reconcile with the evidence of singlet pairing %(primarily 
 from NMR~\cite{brown,newjapan} and from the fact that the in-plane critical field appears to be Pauli-limited~\cite{paulilimit}.

It is always possible that some set of experimental observations are misleading -- perhaps they reflect an extrinsic effect or have an alternative interpretation that has not been considered.  Here, we will take various observations at face value, and attempt  to construct a phenomenological scenario that can  reconcile as many of them as possible.  To do so, we have introduced a single 	``fine -tuned'' assumption -- that Sr$_2$RuO$_4$ happens to be close to a multi-critical point at which  two symmetry-distinct pairing symmetries are degenerate -- in particular, a $d_{x^2-y^2}$ (B$_{1g}$ in group theory nomenclature) and a $g_{xy(x^2-y^2)}$ (A$_{2g}$) state.  %The is the price we have to   pay to  reconcile many  of the seemingly contradictory  observations.  
As we will discuss later in the paper, even with this assumption, there are  some observations that %we have  not successfully reconciled with our scenario, meaning either that our scenario is  wrong or incomplete, or that these remaining experiments have been misinterpreted. \ronny {If we write it like that, we need to give a critical review later on those remainder experiments we cannot integrate into the theory.}
either require a more complete theory to interpret (i.e. the interpretation does not follow simply from general symmetry considerations) or an alternative interpretation of the experiment.

%\ronny{
The article is organized as follows: %} 
% \ronny{In Section~\ref{exp},
{%\color{green} 
we first outline the primary experimental facts we have in mind. Those can be considered the axioms of our phenomenological theory.  In Section~\ref{LG} we set forward the Landau-Ginzburg theory describing a near-degeneracy between a $d$-wave and a $g$-wave SC state, and discuss some features of the resulting phase diagram. In Section~\ref{BCS} we briefly address more features of the phases and phase transitions at a level that would be expected from a BCS-type treatment of the problem;  while these considerations could straightforwardly be extended to include the effects of a  realistic treatment of  the multi-band character of  the SRO fermiology, %we content ourselves with discussing qualitative aspects of the problem that can be suitably addressed from this perspective.
 here we offer a stripped-down discussion that focusses on qualitative aspects of the problem.
  In particular, %this includes 
  we explore the specific heat signatures of the %superconducting transitions we expect in our theory.
  various transitions expected in the theory.
%  in the SC state (for instance, to a state with broken time-reversal symmetry), under conditions in which the two transitions are  somewhat separated.  In Sec. \ref{scenario} we describe at the phenomenological level how this treatment might account for the set of experiments outlined in Sec. \ref{exp} -- no explicit quantitative comparison with experiment is attempted here, %\ronny{but deferred to
%Most importantly, the section further outlines the central subjects of future work related to our phenomenological proposal, which similarly applies to Section~\ref{microscopic} where we sketch some of the microscopic considerations that could prove pivotal to connect to our theory.  
In Section \ref{wrong}, we discuss some of the experimental observations  that are not immediately accounted for by our scenario. This particularly concerns %the evidence from Kerr effect measurements, 
reconciling our proposal with the observed Kerr signal, which requires including considerations of explicit breaking of spatial symmetries %which we argue to match our theory assuming the presence of disorder.
and effects of disorder.  
In both Sections \ref{exp} and \ref{wrong}, we allude to proposed future work that could more seriously take account of specific microscopic aspects of SRO and thus could lead to further avenues for comparing theory with experiment.
 In Section \ref{future},  we conclude by discussing some possible experiments that  could critically test the applicability of our  proposed scenario.}

\section{Experimental ``facts'' in SRO}
\label{exp}

There have been a remarkably large number of extremely careful experiments carried out  to explore the character of the superconducting (SC) state in  Sr$_2$RuO$_4$ (SRO).  We will focus on a subset of these  -  that we will accept as experimental facts with clear implications.  %We defer for now an analysis of others whose interpretation is less clear to us, at least  at  present:
\begin{itemize}
\item{1)} There is a single superconducting (SC) transition with a well-defined $T_c$, and the transition is largely mean-field like.  This inference follows from the fact that the transition as seen in transport and in magnetization is sharp, and the specific heat shows a sharp mean-field-like anomaly at $T_c$.  There is no indication of hysteresis (as  could happen if the SC transition were first order), significant regimes of fluctuation SC above $T_c$, or any glassy feature of the SC transition~\cite{sroreview}.

\item{2)} The superconducting phase is ``unconventional,'' that is the SC order parameter changes sign in such a way that when averaged over the Fermi surface  it is approximately zero.  This inference follows~\cite{mackenzieunconventional} from the extreme sensitivity of $T_c$ to non-magnetic disorder, i.e. $T_c \to 0$ when the elastic mean-free path, $\ell$, approaches the clean limit SC coherence length, $\xi_0$.  (%It also is strongly suggested by  t
The existence of nodal quasiparticles, discussed  below, constitute additional evidence of unconventional pairing.)

\item{3)} To  a  high level of precision, there are gapless nodal quasiparticles along line nodes that are more or less perpendicular to the Ru-O plane, i.e. they extend along the tetragonal c-axis.  This follows from the  low temperature behavior of the specific heat and from the low temperature thermal  conductivity~\cite{thermalconductivity}.  Recent STM studies further corroborate this conclusion~\cite{stm2}.  While the location (in $\vec k$-space) of  the  nodes  is not  entirely clear, both the thermal  conductivity experiments and the STM results are strongly suggestive that the line nodes lie along wave-vectors that are at 45$^o$ to  the Ru-O bond direction, that is along the tetragonal (1,1,0) and (1,-1,0) directions.  %(Some magnetic field dependent  specific heat data are suggestive that the nodes instead lie along the (1,0,0) and (0,1,0) directions %-- we have tentatively disregarded this evidence.as discussed in Sec. \ref{wrong}.)
   If there is any gap at the putative nodal points, it has been estimated~\cite{lupien2001ultrasound,thermalconductivity}  that it  must be less than  $10 \mu \text{eV}$, %\brad{
   or less than %\textit{x}
   3\% of the total gap.

\item{4)} The SC $T_c(\epsilon)$ is a smooth and non-singular function of symmetry breaking (shear) strain, $\epsilon$,  near $\epsilon =0$.
%\footnote{The experiments are, in fact, carried out as a function of applied uniaxial stress, and thus mix both shear and uniform strain.  To simplify the discussion, we will sometimes assume that the principle response seen under these circumstances represents the response to the associated shear component of  the strain, although this assumption certainly needs to be further refined.}
 In the case in which  a %B$_{2g}$ %stress
 {\em stress} is applied along (1,1,0), $T_c$ is a weak linear function of the stress; %\brad{do we want to be clear that by ``linear'' we mean sign-changing, as opposed to the $\|epsilon\|$ dependence you get for the allowed linear coupling}; 
  since by symmetry $T_c$ must be an even function of a purely %linear dependence on
   shear {\em strain}, % is forbidden by symmetry, presumably this 
   the linear stress dependence presumably reflects the fact that, in addition to a $\epsilon_{xy}$ (B$_{2g}$) component, the uniaxial stress produces a significant symmetric component of symmetric  %  strain, 
  strain (with A$_{1g}$ symmetry) $\epsilon_0$.  There is a %shallow, %approximately 
nearly  symmetric variation of $T_c$ in response %B$_{1g}$ 
 to uniaxial stress along (1,0,0). %, a very shallow symmetric dependence symmetric stress dependence is observed. 
  This suggests that, in this case, the induced symmetric %strain
  strain is small, and that what is being observed  is  %the expected even 
  primarily the response to a B$_{1g}$ shear strain, $\epsilon_{x^2-y^2}$.  Local measurements~\cite{kam} (using scanning magnetometry) of $T_c$ vs stress show a quadratic minimum at each location, $T_c(\epsilon) = T_c(0) + \alpha (\epsilon - \bar\epsilon)^2 +  \ldots$, where, however, $T_c(0)$ and $\bar\epsilon$ vary from location to location in the crystal;  the variation of $\bar\epsilon$ %(which has a variance of around 10\%) 
  is, presumably, associated with a random quenched distribution of local strains. Indeed, an interpretation in terms of random strain is strongly indicated based on the fact that some regions of the crystal are seen to have local $T_c$'s
   that are of order 10\% {\em higher} than the zero-strain bulk $T_c$; given the sensitivity of the SC state to impurity scattering~\cite{mackenzieunconventional}, if what was being seen were regions with higher and lower than average impurity concentration, there would be a sharply defined maximum $T_c$ (associated with all sufficiently clean portions of the crystal) and then a distribution of more disordered regions with {\rm reduced} local $T_c$'s.
  %- a form of  disorder that will play a roll in our analysis below.    (
  \footnote{Note that  for large strain, $\epsilon_{x^2-y^2}$, there is a % nearly singular 
  large enhancement of  $T_c$ (from 1.5K to 3.5K) near a critical strain, %$\epsilon_{x^2-y^2} = ***$, 
which is thought to be associated with a Lifshitz transition in the underlying band structure.  The behavior at these larger values of the  strain will not be part of  the present discussion.}
  %\brad{I guess one thing we could look into is whether we can be quantitative with the distribution of strains needed to a account for the distribution in Tc seen by squid. Does some sort of x-ray experiment have the resolution to tell us the ``mosaicity'' of a typical Sr$_2$RuO$_4$ sample? Also, was it clearly pointed out in the squid paper that the distribution of Tc's tended to be \textit{higher} than what is typically thought to be the ``bulk'' Tc of 1.45 K? This is very good evidence that the distribution of Tc isn't just due to disorder -- a fact that would probably escape a good number of readers.}

\item{5)} Time reversal symmetry is spontaneously broken below a temperature $T_{trsb}$ that is (in unstrained  crystals) approximately equal to the SC $T_c$.  This inference follows directly from the  observation of  an anomalous Kerr signal~\cite{kerr}, and from $\mu$Sr experiments~\cite{musr};  it is also consistent with experiments on the geometry dependence of the Josephson relation~\cite{Kidwingira1267},  although beyond supporting the  time-reversal symmetry breaking, this last set of experiments  is  still not fully digested.

\item{5b)} On symmetry grounds, %\ronny{TRSB SC in general %, but chiral superconductivity in particular suggests equilibrium currents to arise} - certainly 
equilibrium currents are possible in a TRSB SC state when not forbidden on the basis of purely spatial symmetries. %, such as at the edges of the sample %and also associated or with various  topological defects (such as domain boundaries)  in the bulk.  
This is particularly true for the case of chiral superconductors  that break both time-reversal and inversion symmetry.  Theoretical estimates of how large these expected currents should be vary greatly depending on details of   the assumed microscopic structure of the state,  but initially suggested that  they should  be detectable by high  sensitivity magnetic scanning microscopy.  However, careful searches~\cite{KAMattheedge} for  the magnetic fields produced by  such currents have failed to detect any such currents, despite sensitivity that is at least two orders of magnitude greater than what would be  required to  detect the currents implied by the most straightforward theoretical analysis (based on an assumed chiral $p_x+i p_y$ state).  

\item{6)} There is a sharp drop in the Knight-shift upon cooling through $T_c$, which  continues to the lowest temperatures and fields that have been probed~\cite{brown,newjapan}.  While it is still necessary to pursue this experiment to lower fields (relative to  $H_{c2}$) and lower $T$ (relative to $T_c$), already the existing observations are strongly suggestive of a spin-singlet order parameter.  The NMR spectrum shows no evidence of a phase transition (within the SC state) as a function of strain;  given this, results at high  strain (where both $H_{c2}$ and $T_c$ are greatly enhanced) can be included in this analysis, making the evidence for singlet pairing considerably stronger.

\item{7)} There is a discontinuity at $T_c$ in the elastic modulus $c_{66}$ associated with the shear B$_{2g}$ ($\epsilon_{xy}$) strain, as inferred from ultrasound experiments~\cite{lupien2002ultrasound,bradspaper}. While the elastic moduli associated with A$_{1g}$ \textit{compressional} strains exhibit discontinuities at all second-order phase transitions, such a discontinuity in a \textit{shear} elastic constant is most straightforwardly associated with a two-component SC order parameter~\cite{ghosh2019single}. This follows from the requirement that a gauge-invariant bilinear can be constructed from the order parameter that transforms as $xy$ under the point group operations -- such a bilinear cannot be constructed for any one-component SC order parameter, but can be constructed for a number of two-component order parameters.  It is also important to note that no discontinuity has been observed in the elastic modulus associated with B$_{1g}$ ($\epsilon_{x^2-y^2}$) strain.

\item{8)} There is %unpublished
new $\mu$-Sr data which shows that for moderate values of stress along the $x$-direction (approximately 1 GPa, which translates to 0.9\% of $\epsilon_{x^2-y^2}$ strain, and 0.3\% of $\epsilon_{x^2+y^2}$ strain~\cite{bradspaper}), that  $T_{tsrb}(\epsilon)=1.0 K$ is substantially below $T_c(\epsilon)=1.4 K$~\cite{newmusr}.  However, the interpretation of this split transition is made somewhat subtle by the observation that in high resolution specific heat measurements, while the critical anomaly at $T_c(\epsilon)$ is still clear and relatively sharp and of the expected magnitude, no thermodynamic signature of any feature associated with a transition at $T_{trsb}(\epsilon)$ has been detected~\cite{li2019high}. 
\end{itemize}

Despite a long history of interpreting results in SRO in terms of a p-wave order parameter (or more technically, an odd parity order parameter) and of the existence of a variety of other experimental results that provide circumstantial support to this idea, we will interpret the new NMR results as ruling this out~\cite{stuartspaper}.
%\ronny{Nodal or close-to-nodal lines along the c-axis would not have been enough to discard the possibility of $p$-wave symmetry, as the preferred $p$-wave state could in principle yield a similar anisotropic gap signature~\cite{Kallin,Ronny}}.%
%\brad{Why do we rule out in-plane d-vectors? My understanding is that the experiment that has been done has only ruled out d-vectors along the c-axis. Is this a similar argument to why we rule out d$_{xz,yz}$ - i.e. it seems weird given the electronic structure of Sr$_2$RuO$_4$.  ---  As we discussed, I think the fact that the NMR sees such a large suppression of the Knight shift rules out p-wave with an in-plane d-vector.  But maybe we should leave this as an assertion and reference it to Stuart Brown's paper.}
%
Focusing on possible singlet (even parity) SC states, consistent with the tetragonal symmetry of the crystal there is one possible symmetry-related two-component state, which corresponds to  $\left\{d_{xz}, d_{yz}\right\}$ (E$_g$) pairing, and four possible single component states, corresponding to $d_{x^2-y^2}$ (B$_{1g}$), $g_{xy(x^2-y^2)}$ (A$_{2g}$), $s$ (A$_{1g}$) and $d_{xy}$ (B$_{2g}$).  

We discard the symmetry-related two-component option (E$_g$)~\cite{PhysRevLett.95.217004,Puetter_2012,Suh} from consideration for two reasons:
%\ronny{Here I cite this recent paper by Dan Agterberg. He claims that chiral $d$-wave does explain the experimental evidence. It might be suited to write two more sentences on his paper her if you all agree. Igor Mazins old paper is more vague and does not commit as strongly as Agterberg does. Brad, Steve, please read that recent arxiv just to be fully educated about it.} %\brad{I guess I might call it the E$_g$ option, since we are about to build a two-component option, and that language might get confusing}:  
 Firstly, the thermal conductivity and STM indicate vertical line nodes while these states have symmetry protected horizontal line nodes.  This is not entirely definitive, since such a state could have accidental vertical line nodes as well, but this seems a rather unnatural occurrence.  %\ronny
 {%\color{green} 
 Secondly, regarding the non-onsite pairing E$_g$ proposal~\cite{PhysRevLett.95.217004}, such a state would have a vanishing pair-wave function for two electrons in the same Ru-O plane, which seems implausible on theoretical grounds,  given the extreme quasi-2D nature of the Fermi surfaces. Regarding the local pairing proposal which involves the multi-orbital nature of the ruthenates~\cite{Puetter_2012,Suh}, it would produce a jump in the elastic modulus for both the $\epsilon_{x^2-y^2}$ and $\epsilon_{xy}$ type strain, whereas only the latter shows a discontinuity in experiments. %Further evidence against this state likely arises from the lack of any 
 Moreover, there is currently no clearly established signatures of the predicted  BdG Fermi surface that arises in such a state. %, and their so far unconfirmed implications for experimental observables such as specific heat~\cite{Irmo}.
 } 

Arguments discriminating among the remaining options are slightly more subtle.  The $d_{x^2-y^2}$ and $g_{xy(x^2-y^2)}$ have symmetry protected vertical line nodes along the (1,1,0) direction, while for any of the other cases, such nodes would necessarily be accidental nodes.  It is not yet clear how precisely the observed line nodes are aligned with this symmetry axis - the more precise the experimental constraints on this, the more solid is the identification of these as the relevant states.  %We will return to the question of which of  the present results survive if our our identification of these  two symmetries as  being relevant to the discussion  proves to be  incorrect.

%The next step in our analysis involves an unnatural assumption.  
Having focused on two distinct 1D representations, the natural next step would be to determine which one is favored.  However, several of the observations summarized above require a two-component order parameter - specifically 5, 7 and 8. %\brad{there is no point 9. Do you mean 8? Does splitting off the magnetic Tc require two components? I guess that makes sese}. 
We are thus driven to  assume that there is an accidental near-degeneracy between $d_{x^2-y^2}$ and $g_{xy(x^2-y^2)}$ pairing, i.e. we assume that SRO happens to be fine-tuned close to  the boundary between a regime in which  one of these is dominant, and the other in which the other dominates.  
%This is, perhaps, not as unnatural an  assumption as it sounds at first.  For example, in the weak coupling limit (where a controlled theoretical analysis is possible)  of the Hubbard model on a square lattice, it was found that as a function of increasing ratio of a nearest-neighbor repulsion, $V$, to an on-site repulsive $U$, there is a transition from the familiar $d_{x^2-y^2}$ phase of the pure Hubbard model to a  $g_{xy(x^2-y^2)}$ for $V/U$ in excess of a critical value, $V \approx ** \rho(E_F) U^2$. Such an analysis has not been carried through for a band-structure of the sort found in SRO, but it is likely that such a transition can arise here as well, especially given the fact that a $d_{x^2-y^2}$ phase arises for a range of conditions with only on-site interactions.  

\section{Landau Ginzburg treatment}
\label{LG}

 Thus, in the following, we will assume that at zero applied strain, SRO happens to be tuned %close
  to a multi-critical point at which these two SC states have (nearly) the same $T_c$.  Depending on other details, the SC phase below $T_c$ can have various patterns of broken symmetry.  This is most  easily explored by considering the 
 %effective potential that enters a Landau-Ginzberg treatment of this problem, 
 Landau-Ginzburg effective free energy density expressed in terms of the  complex  (charge $2e$) scalar  fields, $D$ and $G$, that represent the local value of the pair-fields of the stated symmetries:
 \be
\frac {\cal H}T = % \frac {\kappa_d}2 \left| (-i\vec \nabla -2\vec A) D\right|^2 +  \frac {\kappa_g}2 \left| (-i\vec \nabla -2\vec A) G\right|^2 +  
V_0(D,G) + V_1(G,D) +K+  \ldots
 \ee
 where %$\kappa_a$ is the superfluid stiffness (which properly is a two-tensor quantity), $\vec A$ is the vector potential (which we will set to $\vec 0$ in what follows),
 %
 %  \footnote{The most general ${\cal H}$ allowed by symmetry would also include possible  cross-coupled derivative couplings, such as $\left[(-i\partial_x -2A_x)D\right]^\star \left[(-i\partial_y-2A_y)G\right] + c.c.$.  We don't expect the inclusion of  such terms to change any of the key results of our discussion qualitatively.}
   %
 $V_0$ includes the terms in the effective potential (through quartic order in the fields) in the absence of disorder, external strain (i.e. it assumes tetragonal symmetry), and under the assumption of an exact $Z_2$ symmetry between the $D$ and $G$ orders, while $V_1$ includes to quadratic (leading) order in the fields terms related to breaking of this accidental $Z_2$ symmetry, as well as the effects of disorder and externally applied strain (breaking of tetragonal symmetry), $K$ is a quadratic form in the gauge invariant gradients of the order parameter, $(-i\vec \nabla -2\vec A) D$ and $(-i\vec \nabla -2\vec A) G$, 
 and $\ldots$ refers to higher order terms in powers of the fields and their derivatives.  %Thus
 Specifically,
 \bea
 V_0(D,G) =&& 
 \frac {\alpha_0} 2 \left[\ \left| D\right|^2+\left| G\right|^2\  \right] %- \frac { \alpha_1} 2 \left[ \left| D\right|^2 - \left| G\right |^2 \ \right] 
 + \frac {\gamma_0} 4 \left[\ \left| D\right|^2 + \left| G\right |^2 \ \right]^2 \nonumber \\
+&& 
\frac {\gamma_1} 4 \left[\ \left| D\right|^2 - \left| G\right |^2\ \right]^2 + \frac {\gamma_2} 4 \left[\ D^\star G+ G^\star D \right]^2 %- \frac {\gamma_3} 4 \left[\ D^\star G- G^\star D \right]^2
 \eea
 %
% \brad{are the $\gamma_0$ and $\gamma_1$ terms correct? I think this are supposed to be the quartic terms, yes? Then the G's are missing a square.} 
%
and
 \be
V_1(D,G) = -\frac{h_0(\vec r)}2  \left[\ \left| D\right|^2+\left| G\right|^2\  \right]-\frac{h_1(\vec r)}2  \left[\ \left| D\right|^2-\left| G\right|^2\  \right] \ .
  -\frac {h_2(\vec r)}2   \left[\ D^\star G+ G^\star D \right]
 \ee
 Note that $V_0$ has an enlarged $SO(4)$ symmetry for the special case in which $\gamma_1=\gamma_2=0$.
 
 Here, the principal $T$ dependence  is incorporated in a $T$ dependence of $\alpha_0$, such that it changes from positive to negative as $T$ drops from above to below $T_c$.  The multicritical point that serves as the focus point for our  analysis arises when $ %\alpha_1 =0
 h_1=0$;  otherwise, the mean value of $ %\alpha_1
 h_1$ encodes the preference for one or the other form of pairing, with $d$-wave pairing favored for $ %\alpha
 h_1 >0$.  Below $T_c$, the nature of  the  SC state  is determined by the signs and relative magnitudes of  the quartic terms.  To simplify the discussion, %we will assume that  the approximate symmetry between the two components extends past lowest order, which is to say that 
 we will consider the case in which %the fully SO(4) symmetric terms, 
  $\gamma_0 $ is positive and larger in magnitude than the remaining terms. %;  in this case, we can always normalize  the SC fields so that $\gamma_0=1$.
In the uniform SC state when $h_1=0$, the preferred form of ordering is determined by the  %$\gamma$'s is smallest. 
relative sign and magnitude of $\gamma_1$ and $\gamma_2$. 
For $\gamma_1<0$ and $\gamma_2 > \gamma_1$  the SC state is either pure  $d$ or pure  $g$, and thus preserves all spatial symmetries as well as time-reversal symmetry. 
The case that will be of primary interest here is that in which $\gamma_1>0$ and $\gamma_2 >0$, so the SC state breaks both time-reversal symmetry as well as some lattice symmetries;  we refer to this as a TRSB SC, and it corresponds to $d_{x^2-y^2}\pm i g_{xy(x^2-y^2)}$ pairing. %, again depending on the relative phase of the two  components.
Finally, if $\gamma_1 > \gamma_2$ and $\gamma_2 < 0$ the SC state breaks various lattice symmetries but preserves time-reversal symmetry;  this is a  nematic superconductor, which has two possible symmetry related states  $d_{x^2-y^2}\pm g_{xy(x^2-y^2)}$, depending on the relative sign of the two components. 

%is the smallest, then the SC state is either pure  $d$ or pure  $g$, and thus preserves all spatial symmetries as well as time-reversal symmetry.  If $\gamma_2$ is the smallest, then the SC state breaks various lattice symmetries but preserves time-reversal symmetry;  this is a  nematic superconductor, which has two possible symmetry related states  $d\pm g$, depending on the relative sign of the two components. Finally, if $\gamma_3$ is the smallest, the SC state breaks both time-reversal symmetry as well as some lattice symmetries;  we refer to this as a TRSB SC, and it corresponds to $d\pm i g$ pairing. %, again depending on the relative phase of the two  components.
 
\begin{figure}
 	\centering
 	\includegraphics[width=1\columnwidth]{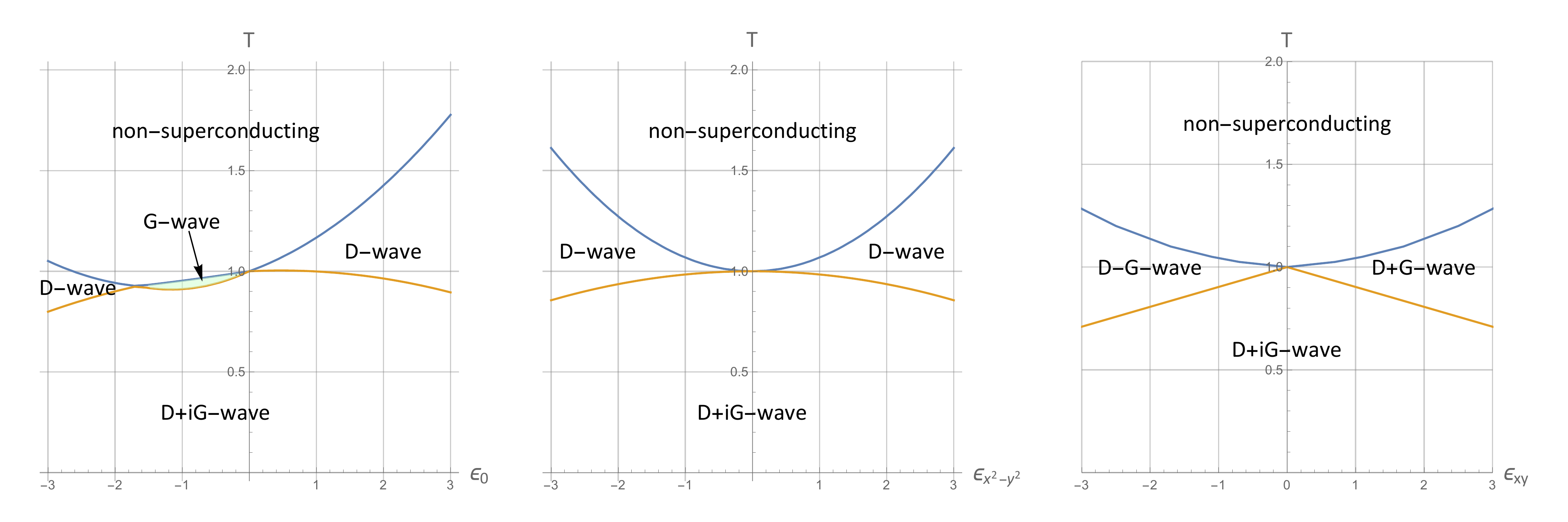}
		\caption{Schematic phase diagrams as a function of $T$ and various components of the strain.  In keeping with the present scenario, we assume that the D and G wave orders are essentially degenerate at zero strain.  All panels were simulated with LG parameters $\gamma_0=2,\gamma_1=1,\gamma_2=0.5$. Panel a) is for symmetric (A$_{1g}$) strain. Note that the shape and even the topology of the phase diagram depends on the relative sign and magnitude of 
		%$\Lambda_0^0$, $\Lambda_1^0$, $\Lambda_0^{0,0}$, and $\Lambda_1^{0,0}$, which % in our case 
		the various couplings $\Lambda_j^a$ and $\Lambda_j^{ab}$;
		here, for purposes of illustration, %were chosen to be 
		we took $\Lambda_0^0=-.086,\Lambda_1^0=.035,\Lambda_0^{0,0}=-.025$, and $\Lambda_1^{0,0}=.021$.  Panel b) is for B$_{1g}$ shear strain, %Now,  the shape of the phase diagram depends on the relative sign and magnitude of $\Lambda_0^{a,a}$ and  $\Lambda_1^{a,a}$,with $a=(x^2-y^2)$, which were chosen to be
		where the explicitly shown phase boundaries correspond to  $\Lambda_0^{a,a}=-.04$ and $\Lambda_1^{a,a}=.028$  with $a=(x^2-y^2)$. Panel c) is for B$_{2g}$ shear strain. Here,  to linear order in the strain, the shape of the phase diagram  is %independent of details, but for completeness, we include the used parameter values, 
		insenstive to the assumed parameters, here chosen to be $\Lambda_0^{b,b}=-.02,\Lambda_1^{b,b}=.01$, and $\Lambda_2^b=.014$ where $b=xy$.}
	 	\label{phasediagrams}
 \end{figure}

% The significance of the different symmetries of these various states is made manifest when one considers how they couple to externally applied strain or extrinsic disorder. In the  absence of an  external magnetic field or any other source of explicit time-reversal symmetry breaking, only the first two forms of order couple directly to the disorder potential to lowest order.  Specifically, to quadratic order in the fields, %the effect of quenched randomness
% these effects can be incorporated into a Landau-Ginzburg effective field theory by adding to the effective potential
% \be
%  \delta V(D,G) = -\frac{h_0(\vec r)}2  \left[\ \left| D\right|^2+\left| G\right|^2\  \right]-\frac{h_1(\vec r)}2  \left[\ \left| D\right|^2-\left| G\right|^2\  \right]
 % -\frac {h_2(\vec r)}2   \left[\ D^\star G+ G^\star D \right]
% \ee
Turning to the terms that appear in $V_1$,
  $h_0$ has the interpretation as a %random
   local $T_c$, $h_1$ as %a random 
   the local deviation from multicritcality (i.e. a local preference for $d$ or $g$), and $h_2$ has the symmetry of a random local strain, $\epsilon_{xy}$.  Importantly, such a local strain favors one orientation of the nematic order over  the other. %\brad{This is where I get a bit unfamiliar with what is allowed and what is not - I would appreciate a short tutorial on introducing disorder sometime over Skype}.
 %Note  that the same symmetry considerations  govern the coupling of  the SC order to uniform (externally applied) strains.  % - i.e. this  coupling to the nematic component has the correct  symmetry to account for the elastic constant anomaly already mentioned  above.
 %
 In the absence of an externally applied strain, the configuration (or spatial) average of all the random fields can be assumed to vanish, $\overline{h_j(\vec r) } = 0$, but in the presence of applied strain, 
 \be
 \overline{h_j(\vec r) } =  \Lambda_j^{a}\ \epsilon_a+ \Lambda_j^{a,b}\epsilon_a\epsilon_b  + \ldots, 
 \ee
 where a sum over $a$ is implicit, $\Lambda_j^a$ and $\Lambda_j^{ab}$ are appropriate coupling constants, the indices $a$ and $b$ (which are implicitly summed over) label the components of the strain  tensor - which we decompose according to  symmetry  as $a=s$ for A$_{1g}$ (although in a tetragonal  system, there are properly two distinct components of A$_{1g}$ strain), $a=(x^2-y^2)$ for B$_{1g}$, and $a=xy$ for B$_{2g}$,  and $\ldots$ indicates higher order terms in powers of the  strain.   All components of $\Lambda_j^a=0$ by symmetry other than $\Lambda_0^s$, $\Lambda_1^s$, and $\Lambda_2^{xy}$. %;  uniform strain acts to shift the mean $T_c$, but also to produce a splitting between $D$  and $G$ orders.  
 Similarly, all terms in $\Lambda_j^{ab}=0$ other than the diagonal terms, $\Lambda_0^{a,a}$ and $\Lambda_1^{a,a}$ (for any $a$), and the off-diagonal term, $\Lambda_2^{s,xy}= \Lambda_2^{xy,s}$.  In other words, there generically is a shift of the mean $T_c$ and of the splitting between the $d$ and $g$  wave order that is linear in  the isotropic  strain and quadratic in each component of  the shear strain.  However, the degeneracy between $d_{x^2-y^2}+g_{xy(x^2-y^2)}$ and $d_{x^2-y^2}-g_{xy(x^2-y^2)}$ %requires 
 is lifted only in the presence of non-zero B$_{2g}$  strain.
%  $\ldots$ indicates higher order terms in powers of the  strain. % \brad{This paragraph as well, I think I just need you to go over the logic with me}.
  %{\color{green} 
   The detailed shape and  even topology of the resulting phase diagrams in the $T$-strain plane depend on the relative magnitudes of various parameters in the Landau-Ginzburg free energy;  representative phase diagrams as a function of strain (assuming degeneracy between d and g at zero strain) are shown in Fig. \ref{phasediagrams}
  
  When we come  to discuss  disorder, we will  define the root mean squared disorder strengths, $\sigma_j$, as
  \be
\nu \int d\vec r \   \overline{\left[h_i(\vec r)  - \overline{h_i}\right]\ \left[h_j(\vec 0)   -\overline{h_j}\right] }\equiv \delta_{i,j} \sigma_j^2
\label{variance}
\ee
(where $\nu$ is the volume of a unit cell).
Because the most relevant disorder is likely that  associated with inhomogeneous strain, it may be important to consider the range of the disorder correlations, as well.

\section{%Microscopic (BCS) considerations
Quasiparticle properties;  BCS considerations}
 \label{BCS}
 
 %Note that w
 While we are free to normalize the order parameter fields that appear in the Landau-Ginzburg effective field theory in any way we like, the same freedom does not pertain when we relate these fields to the gap parameter that governs the quasi-particle spectrum.  Interpreting the various  possible states through the lense of the  BdG equations of BCS mean-field theory, we relate  the gap  parameter  to the order parameter fields according to
 \be
 \Delta(\vec k) =  e^{i\theta}\  \left[D\ F_d(\vec k)\   + e^{i\phi}G \ F_g(\vec k) \right]
 \ee
 where $\theta$ is the overall SC phase, $\phi$ is the relative phase between the d and $g$-wave components, and 
 where we normalize the  real dimensionless gap functions $F_a(\vec k)$ %such that 
 so that ${\rm Max}[F_a(\vec k)] = 1$;  thus $D$ and $G$ are, respectively, the maximum gap in the $d$-wave and $g$-wave state, respectively.
 % \be
 %\nu  \int \frac {d\vec k} {(2\pi)^3} \ \left| F_a(\vec k) \right|^2 = 1
 %\ee
 $F_a$ can have an arbitrarily complicated $\vec  k$ dependence   (depending on microscopic details), so long as they transform in the desired fashion under the crystalline symmetries.  
 % In  principle, %all    the form of the gap functions, $F_a$,  can depend on $T$ and strain as well, but 
 %   In particular, $\gamma$ is a dimensionless parameter which characterizes the relative strength of the $d$ and $g$ contributions to  the gap in circumstances in which  both are fully developed.  
 %
 In keeping with the spirit of the Landau-Ginzburg approach, at least in a range of $T$ near to $T_c$, we can imagine that all the singular $T$ and strain dependences in the problem are inherited from those of the order parameter fields, $D$ and $G$. 
 %
% For simplicity, when we perform estimates that require an explicit form, we will assume the simplest forms consistent with symmetry: $F_d =(1/2)\left [\cos(k_x) -\cos(k_y)\right]$ while $F_g =(3\sqrt{3}/4) \sin(k_x)\sin(k_y)\left[\cos(k_x) -\cos(k_y)\right]$. \ronny{Steve, these are the shortest ranged harmonics for d and g. I bet this won't be what we get from weak coupling, we probably get an important admixture of longer ranged harmonics. Indeed, for FRG, these harmonics are most likely the by far most dominant one. F$_d$ and F$_g$ are exteremely important as soon as we calculate something. Is this what enters Andrews calculations? Are we then just plotting $\Delta=\vert F_d+iF_g \vert$ in a figure or do you want a more sophisticated way to calculate a first guess for the SC gap on the Fermi level? I cannot see the figure 1 you are referring to later.}
%
 From here, as usual, it follows (still assuming the validity of BCS mean-field theory) that the quasiparticle spectrum is
 \be
 E(\vec k) = \sqrt{ [\varepsilon(\vec k) - \mu]^2 + |\Delta(\vec k)|^2 }
 \ee
 where $\varepsilon(\vec k)$ is the normal state (Fermi liquid) dispersion.
 
 \begin{figure}
 	\centering
 	\includegraphics[width=0.7
 \columnwidth]{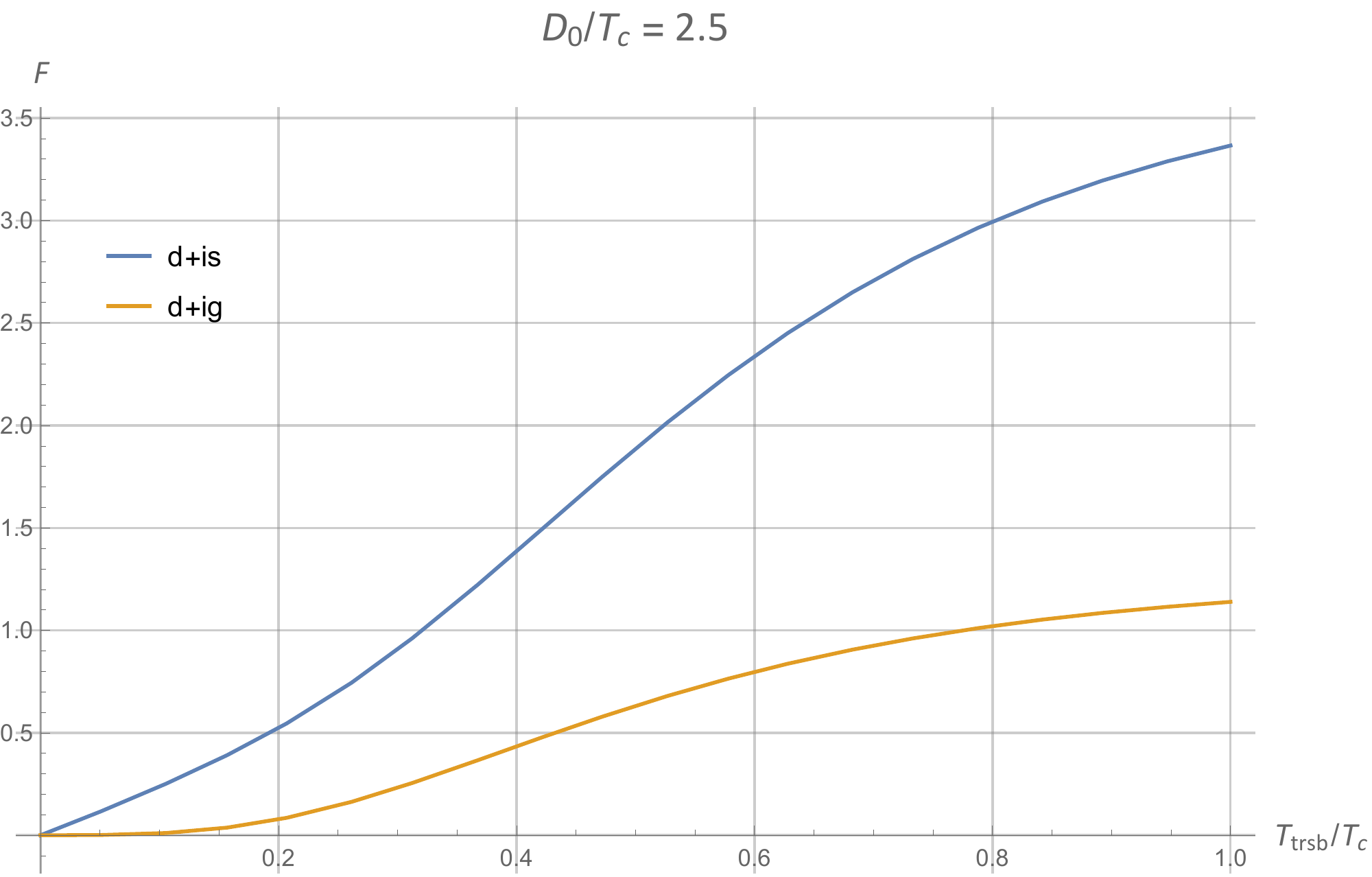}
 	\caption{Dimensionless magnitude of the specific heat jump at $T_{trsb}$ -- ${\cal F} (T_{trsb}/T_c)$ from Eq. \ref{jump}, taking  $D_0/T_c=2.5$.  The lower curve is for the case in which the transition is from a $d$-wave to a $d_{x^2-y^2}+i g_{xy(x^2-y^2)}$-wave SC;  the upper curve is for a $d$-wave to a $d_{x^2-y^2}+i s$-wave SC.}
 	\label{fig:Fig1}
 \end{figure}
 At the level of BCS mean-field theory, once the values of $D(T)$ and $G(T)$ are determined, a host of dynamic and thermodynamic quantities can be computed in the usual way.  One particularly important quantity is the jump in the specific heat at the various transitions.  Consider, for example, the case in which $\overline{h_1} >0$ and $\gamma_1$, $\gamma_2> 0$, so that upon lowering the temperature there is first a transition at $T_c$ to a pure $d$-wave SC state, and then at a lower temperature, $T_{trsb} < T_c$, a transition to a $d_{x^2-y^2}+i g_{xy(x^2-y^2)}$ state.  %We assume the usual 
 If we assume the simplest possible mean-field dependence of the gaps
 \be
 D(T) = D_0 \sqrt{1 - \frac{T}{T_c}} \ {\rm for}  \ T<T_c
 \ee
 \be
 G(T) = G_0 \sqrt{1 - \frac{T}{T_{trsb}}} \ {\rm for}  \ T<T_{trsb} < T_c \ ,
 \ee
% In terms of these, t
the specific heat jump at $T_c$ can be expressed as
 \be
 \frac {\Delta C_{sc}} {T_c} =\left(\frac{D_0}{T_c}\right)^2 %\overline
 \left[\left[{\left(-\frac{\partial f}{\partial E} \left|F_d\right|^2\right)}\right] \right] =   \left(\frac{D_0}{T_c} \right)^2\ \rho(E_F) \  \langle \left| F_d\right|^2 \rangle \     % \rho_F
 \ee
 and the specific heat jump at $T_{trsb}$ as
 \be
 \frac {\Delta C_{trsb}} {T_{trsb}} = \left(\frac{G_0}{T_{trsb}}\right)^2 \ \rho(E_F)\ \langle \left|F_g\right|^2 A \left(z \right) \rangle   = \frac {\Delta C_{sc}} {T_c} \left(\frac {G_0}{T_{trsb}}\right)^2\ \left( \frac{T_c}{D_0} \right)^2    {\cal F}\left(\frac {T_{trsb}}{T_c}, \frac {D_0}{T_c}\right)
 \label{jump}
 \ee
 where $\langle f \rangle$ and $%\bar
 \left[\left[ f\right]\right]$ signify, respectively, the average of $f$  over the Fermi surface and over the Brillouin zone,
% {\color{green} There is a problem with the definition of A}
% \be
 %\langle f\rangle = \frac{\overline{f \delta(\xi_k)}}{\rho_F}, \qquad \bar{f} = \int \frac{d^2 k}{(2\pi)^2} f(k)
% \ee
 \be
 A(z)=\int_0^\infty \frac{dx}{\cosh^2{\sqrt{x^2+z^2}}},  \qquad z =\frac{F_d}{2} \frac{D_0}{T_{trsb}}\sqrt{1-\frac{T_{trsb}}{T_c}} \ ,
 \ee
% and 
 $E_d(\vec k)$ is the dispersion in the pure $d$-wave SC state, and  ${\cal F}$ is a dimensionless function defined  implicitly in Eq. \ref{jump}.

 In Fig. \ref{fig:Fig1} we have computed ${\cal F}$ as a function of $T_{trsb}/T_c$ for a band-structure corresponding to nearest-neighbor hopping on a square lattice with a mean electron density $n=1.38$, per site 
 and assumed
$F_a$ have   the simplest forms consistent with symmetry, $F_d =(1/2)\left [\cos(k_x) -\cos(k_y)\right]$ and $F_g =(3\sqrt{3}/4) \sin(k_x)\sin(k_y)\left[\cos(k_x) -\cos(k_y)\right]$,
  and a typical mean-field value of  {%\color{green} 
  $D_0/T_c=2.5$.  For comparison, we have performed the same calculation foo the case in which the second order parameter is a simple s-wave, $F_g \to F_s=1$.
  It is striking that the specific heat jump is much smaller for the transition to the d+ig state, and especially so the smaller $T_{trsb}/T_c$.}
  %It is striking how rapidly the specific heat jump at the lower transition becomes small as soon as the two transitions are substantially separated.  
  There are several contributing factors to this.  Firstly, because the $d$-wave gap grows so singularly with decreasing temperature below $T_c$, the total remaining entropy in the system is rapidly reduced relative to that in the normal state -- this effect would apply for any ``second'' transition in the SC state.  However, the effect is particularly dramatic due to the fact that the $g$-wave gap  also has nodes in the one region of the Fermi surface that remains ungapped (nodal) in the $d$-wave SC state.  Thus, the density of thermally excited quasiparticles is only %very 
  slightly (further) reduced upon entering the d+ig - TRSB phase.

 \section{A plausible scenario}
 \label{scenario}
 
The set of experimental observations that we have enumerated above have, for some time, seemed difficult to reconcile with each other.  With the assumption that SRO is tuned  close to the conjectured multi-critical  point, much of this seeming difficulty is readily eliminated. % even without considering the effects of quenched disorder.  
 For ${\overline h_1}=0$, there is a single SC transition, and the  SC state  is unconventional, in keeping with points 1 and 2 . The  SC order is even parity, spin-singlet pairing, consistent with point 6.   Because both the  $d$ and the $g$ wave components of the order  parameter are odd under any spatial transformation that exchanges $k_x$ and $k_y$, consistent with point 3, the existence of nodal lines in the quasi-particle spectrum along the (1,1,0) direction is  protected by symmetry, even in the TRSB SC phase;  this is distinct from the case for any other order parameter symmetries that have been considered to date. 
 %\ronny{ From the viewpiont of topological superconductivity, even ignoring the fact of symmetry-protected nodel lines, the SC order of a d+ig state preserves parity and hence is non-chiral, i.e. would not feature a non-zero Chern number of the associated Bogoliubov bands.}  
% Finally, %as a consequence of the assumption $\alpha_1=0$, 
Importantly,
 at the assumed multicritical point,
 the $d$ and $g$-wave components of the order parameters behave like a two-component order parameter with precisely the requisite relative symmetries to account for point 7. Specifically, the bilinear formed of B$_{1g}$ and A$_{2g}$ order parameters is of the B$_{2g}$ representation, thus producing a discontinuity in $c_{66}$ at $T_c$. This scenario also provides a natural explanation for why no discontinuity is seen in $(c_{11}-c_{12})/2$---the B$_{1g}$ elastic modulus---since such a bilinear cannot be formed. This is in contrast with other two-component superconducting order parameters, such as d$_{xz,yz}$ or p$_{x,y}$ states, where a discontinuity in $(c_{11}-c_{12})/2$ is expected in general, and the lack of experimental evidence for such a discontinuity would have to be explained away in some other way.   %by invoking, for example, large attenuation in this strain channel. \brad{I added a few more specifics, and then also a discussion of how this d+ig state actually does \textit{better} than the other two-component states, in some respect, because it doesn't produce a jump in B1g. Not sure you want all this here, but it may belong somewhere.}
 %For the same reason, no non-analytic dependence of $T_c$ or $T_{trsb}$ on B$_{1g}$ strain is expected;  $T_c$  and $T_{trsb}$ should vary in proportion to $\epsilon_{x^2-y^2}$ for small strain, as seen empirically.
 
 Since $T_c$ appears to be most strongly sensitive to   $B_{1g}$  strain, it is important to stress that % in the present  case, even in the clean limit, the application of $B_{1g}$  strain, $\epsilon_{x^2-y^2}$, 
 %\brad{This is the B1g strain, not the B2g strain. I didn't correct it because I'm not 100\% sure which you are referring to},
%even in the clean limit, no % is not expected to cause  
no  singular non-analytic changes in either $T_c$ or $T_{trsb}$ as a function of  $\epsilon_{x^2-y^2}$ are expected.   %We conjecture that t
 The principal effects of strain of  this symmetry are analytic (i.e. quadratic) strain dependences of % $\alpha_0$ and $\alpha_1$. 
 $\overline{h_0}$ and $\overline{h_1}$.   %On general grounds, there
 There is no general  reason for one sign of  the effect, % or the other, 
 but both empirically, and based on microscopic considerations related to the proximity of one band to a  Lifshitz transition point, it is clear that %$\alpha
 $h_0$ (and hence $T_c$) must be a strongly  increasing function of $[\epsilon_{x^2-y^2}]^2$.  %Of more qualitative significance, we would  expect that any form of strain would tune the system away from the multicritical condition. %, $\alpha_1=0$.  
 %Again, the sign of the effect cannot be deduced on any general grounds, but rather %follows from microscopic considerations.  
 It is also reasonable to assume on microscopic grounds that $h_1$ is an increasing function of $[\epsilon_{x^2-y^2}]^2$,  since the fact that the $g$-wave gap function vanishes at the van-Hove points implies that
 proximity to the %van-Hove point
 Lifshitz transition likely favors $d$-wave over $g$-wave pairing. %\brad{why?},   
 %Given that various (weak coupling) microscopic calculations have already predicted a propensity for $d$-wave pairing, we conjecture that the more likely result is that strain favors the $d$-wave state, i.e. %$\alpha $ \overline{h_1} \sim + [\epsilon_{x^2-y^2}]^2$.  
%;  assuming the specific strain dependences outlined above, this leads to a strain dependence 
% $T_c(\epsilon_{x^2-y^2})-T_c(0) \sim + [\epsilon_{x^2-y^2}]^2$, below which the system would form
Consequently, as shown in Fig.~\ref{phasediagrams}, we expect that $T_c - T_{trsb} \sim + [\epsilon_{x^2-y^2}]^2$ and that   for $T_c > T>T_{trsb}$, there is a  $d$-wave-like SC %(or more precisely, since  the  tetragonal symmetry  is explicitly broken, a d+s-wave SC 
%\brad{this d+s langauge really confuses a lot of people - they think it means two-component (and it has been suggested to me that I look for the jump in say YBCO, which is d+s.) This is of course just bad notation - $x^2$ and $y^2$ both transform as A1g in D2h, so you are just getting $a x^2$ + $b y^2$, which looks like a bit of $x^2 + y^2$ and a bit of $x^2-y^2$. You know this of course, but is it important to be explicit? Maybe not?} ),
 followed by  a lower $T$  transition to a  TRSB SC with $d_{x^2-y^2}+i g_{xy(x^2-y^2)}$-wave-like order. %with  $T_c(\epsilon_{x^2-y^2}) - T_{trsb}(\epsilon_{x^2-y^2}) 
 (Strictly speaking, in the presence of shear strain, the SC state should be classified by the representations of the orthorhombic point group which is why we refer to the states as ``$d$-wave like'' and ``$g$-wave like.'')
%where $T_c - T_{trsb} \sim + [\epsilon_{x^2-y^2}]^2$.  %Note that both these effects would be  expected to be impervious to weak disorder \brad{what exactly do we mean by disorder here? Strain, or high-Q scattering?}, and that both of these observations 
%(The issue of why the second transition has such a weak signature in the specific heat will require further study:  The fact that the gap structure does not actually change all that much in going from the d to the d+iG state, and that in fact the nodal structure is entirely unchanged certainly could play a role in this. Moreover, if  -  as we discuss  below - the TRSB occurs primarily on relatively dilute domain walls, this could further reduce the thermodynamic signature of the second transition.)
This all is broadly  consistent with point 8~\footnote{The lack of an observable  signature of $T_{trsb}$ in the specific heat is a quantitative issue that will need further investigation.  However, as we have shown, the fact that the nodal structure of the $d_{x^2-y^2}+i g_{xy(x^2-y^2)}$ SC is the same as for the pure $d$-wave SC leads to a significant reduction in the expected size of the specific heat jump at $T_{trsb}$.  Moreover,  effects of random strain can be expected to broaden and further reduce the magnitude of any specific heat signature of $T_{trsb}$.}.

 The TRSB in the SC state discussed as point 5 %would be
 is  consistent with  the present scenario %were we to assume that $\gamma_3 < \gamma_1$, $\gamma_2$  (and, of course, $\gamma_3 < 1$). 
 so long as we assume all $\gamma_j>0$. Precisely at the point at which the two orders are degenerate, the onset of TRSB would occur simultaneously with $T_c$, consistent with present observations.  However, as is exhibited in the various schematic phase diagrams in Fig.~\ref{phasediagrams}, any strain is expected to drive the system away from this fine-tuned condition, resulting in a phase diagram in which $T_c > T_{trsb}$, consistent with point 8.  
 %Indeed, as we have shown,
  The fact that one set of the $g$-wave nodes coincides with the $d$-wave nodes means that the specific heat jump at the transition from a pure $d$-wave SC to a $d_{x^2-y^2}+i g_{xy(x^2-y^2)}$ state is significantly reduced, %possibly accounting
  which may partially account  for the fact that no signature of $T_{trsb}$ has yet been observed in specific heat measurements.  (Note, in a regime in which the higher $T$ transition is to a pure $g$-wave SC, one would expect a stronger thermodynamic signature of the lower temperature transition to the $d_{x^2-y^2}+i g_{xy(x^2-y^2)}$ state since one set of $g$-wave nodes would be gapped below $T_{trsb}$.
  
% How large the edge currents associated with a $$d_{x^2-y^2}+i g_{xy(x^2-y^2)}$$ SC has not (to the best of our knowledge) been computed theoretically, so it is not absolutely clear to what extent this is consistent with point 5b.  However, t
There is good reason to expect small or even vanishing equilibrium currents in this state.   Since the SC order of a $d_{x^2-y^2}+i g_{xy(x^2-y^2)}$ state preserves parity, it is non-chiral, i.e..the associated Bogoliubov bands   have zero Chern number.  Thus there is no topological reason to expect edge currents.  
% Indeed,
Moreover, the fact that reflection through a mirror plane perpendicular to the (110) direction is respected  even in the $d_{x^2-y^2}+i g_{xy(x^2-y^2)}$ state is  sufficient to establish that edge currents must vanish at any surface that respects  this symmetry.    
%\ronny{Steve, a principal question: How offensively should we discuss the issue of a Kerr singal for a parity preserving superconductor? So far I believe we say nothing wrong and we hint at the right things, but would think it is better to address this issue more directly?}

However, despite these successes,  one is hard-pressed in the context of the present scenario to simultaneously account for a portion of the observations in point 4 (no non-analytic $\epsilon_{xy}$ strain dependence of $T_c$ in the limit of vanishing   stain).
% and point 8 (the appearance of a distinct $T_{trsb} < T_c$ under B$_{1g}$ strain, with no associated (detectable)  singularity in the specific  heat).  
To account for %these additional observations
this  in a consistent fashion we invoke the  effect of weak disorder that is equivalent to random local strains.  Note that while it is an established fact that the relevant materials are, to an extraordinary level of approximation, nearly perfectly crystalline,  with normal state mean-free paths in excess of 1 $\mu$m  (point 2), long-wave-length disorder (i.e. strain disorder) does not make much contribution to determining the mean free path, and is not expected to lead to  a substantial reduction of $T_c$, even in an unconventional  SC.  Indeed, as discussed in point 4, there is  some direct experimental evidence of the existence of the presence of %(small)
 random strain~\cite{kam}.

There are several  different  scenarios  that can (at least roughly) reconcile the remaining observations by incorporate  the  effects  of  weak,  random strain -- that is to say non-zero $\sigma_j$ as defined in Eq.~\ref{variance}.  We plan to address this issue in a forthcoming work.  In general, for the same reasons discussed in Ref. \onlinecite{yue} in the context of an assumed two-component p-wave order parameter, such disorder  rounds out any non-analytic cusp-like strain dependence of $T_c$.  The easiest limit in which to illustrate this is one in which the random strains are very long-range correlated, so one can think of the system as consisting  of essentially macroscopic regions with different strains.  In any region with a non-zero value $h_1$, the accidental degeneracy between d and g is lifted, so even a local measurement of $T_c$ versus strain would see analytic dependence.  In regions with approximately zero $h_1$, but non-zero $h_2$,  a cusp would be expected in the $\epsilon_{xy}$ dependence of $T_c$ measured locally.  However, even in this case, the global onset of superconductivity would be determined by the point at which superconducting regions percolate, which would have no such cusp~\footnote{The scanning SQUID measurements of Ref. \onlinecite{kam} establish that even locally, the dependence of $T_c$ on $\epsilon_{x^2-y^2}$ strain is quadratic, such experiments have not been carried out for $\epsilon_{xy}$ strain.}.
One would also expect strain disorder to further reduce both the magnitude and the sharpness of any thermodynamic signature of the transition to the TRSB state in the case in which $T_c$ is measurably larger than $T_{trsb}$.
 
% Whether the present scenario is consistent with point 5b requires further study.  However, it is encouraging that previous studies have shown that in cases in which the gap parameter has multiple sign changes along the Fermi surface, the expected edge currents are greatly reduced in magnitude.
 
% \ronny{ From the viewpiont of topological superconductivity, even ignoring the fact of symmetry-protected nodel lines, the SC order of a $d_{x^2-y^2}+i g_{xy(x^2-y^2)}$ state preserves parity and hence is non-chiral, i.e. would not feature a non-zero Chern number of the associated Bogoliubov bands.}  
 
 \section{Aspects of a microscopic picture}
 \label{microscopic}

%\ronny{Steve, just a suggestion: Would you consider keeping your main idea with the nearest neighbor coupling into the part above where you define the short range harmonics for F$_g$ and F$_d$, and get rid of this section? It is too long to be ignored and too short to understand it.}
  
Because the normal state of SRO is a good  Fermi liquid, and  because the SC phase transition appears to be mean-field like, it makes  sense to think about the mechanism of SC from a weak-coupling BCS perspective. % \ronny{
 Various  studies using an  asymptotically exact weak-coupling  approach  to the problem taking into account only on-site interactions have found  that  the SRO band-structure is conducive to SC states with %preferably 
p-wave symmetry 
and %subleading
  d$_{x^2-y^2}$-wave symmetry~\cite{raghuandme,PhysRevLett.105.136401,Steppkeeaaf9398}.  %\brad{So I have no way of evaluating what was done in those studies - how much of the evidence for p-wave was just playing around with parameters because people thought they should get p-wave? Is there such a thing as a completely blind estimate of what the weak-coupling ground state symmetry is, or is such a calculation just way too naive?}.  
Similar conclusions have been  drawn on the basis of a random phase approximation ansatz~\cite{hong} and multi-orbital functional renormalization group (FRG) studies which %are somewhat less well controlled, but which 
 better account for the strong  antiferromagnetic fluctuations seen in neutron scattering  experiments on SRO~\cite{Ronny,PhysRevLett.122.027002}.
  In %both 
 all cases, the pair-wave-function approximately vanishes on site, thus avoiding the effects of a repulsive on-site Hubbard $U$.  Given  the  uncertainty concerning the correct relative  magnitudes of the various interactions, and  the intrinsic issues concerning the applicability of such approaches to real materials (where interactions are never weak), %the  accuracy of such weak  coupling approaches  in  real  problems (with  intermediate strength interactions \brad{The paper (And this paragraph) mention weak-coupling a few times, but this is talking about intermediate-strength. Are these referring to two different things? Can something be intermediate strength but still weak coupling? Intermediate seems to imply that interactions are of the scale of the bandwidth, whereas weak coupling they would be small (and strong coupling implies they are bigger than the single-particle bandwidth)?} ), 
 even thought the p-wave instability is often found to be dominant,
 it would be difficult to argue strongly on purely theoretical grounds which of these is preferred.  However, as far as we know, no previous study of SRO has found indication of a  significant tendency to  pairing in a g-wave channel.

However, in a related study, the effect of a repulsive nearest-neighbor repulsion, $V$, on the character of the weak coupling SC  for the  Hubbard model on a   square lattice  was analyzed~\cite{PhysRevB.85.024516}.  In  the range of doping in which  the dominant pairing instability  is in the  $d_{(x^2-y^2)}$ channel at  $V=0$, there  is a transition to a g-wave state for  $V$  greater than a  critical  value,  $V_c \sim U^2  \rho(E_F)$. This is natural  in   the  sense that the pair wave-function  in the  g-wave state  not  only  vanishes on-site, but  also on  any pair  of  nearest-neighbor sites.  Thus, for $V=V_c$, precisely the sort of accidental degeneracy we have invoked arises.  (We are currently in the process of extending  this analysis  to the  case of the SRO bandstructure~\cite{inprep}.) Alternatively, even though probably more contrived than the latter microscopic reasoning, a proximity of $d_{(x^2-y^2)}$-wave and $g$-wave pairing propensity had been found in an effective two-dimensional electronic model inspired by the ruthenates which was assumed to be subject to three-dimensional phonon coupling~\cite{PhysRevB.74.184503}.
 
 %This is, perhaps, not as unnatural an  assumption as it sounds at first.  For example, in the weak coupling limit (where a controlled theoretical analysis is possible)  of the Hubbard model on a square lattice, it was found that as a function of increasing ratio of a nearest-neighbor repulsion, $V$, to an on-site repulsive $U$, there is a transition from the familiar $d_{x^2-y^2}$ phase of the pure Hubbard model to a  $g_{xy(x^2-y^2)}$ for $V/U$ in excess of a critical value, $V \approx ** \rho(E_F) U^2$. Such an analysis has not been carried through for a band-structure of the sort found in SRO, but it is likely that such a transition can arise here as well, especially given the fact that a $d_{x^2-y^2}$ phase arises for a range of conditions with only on-site interactions.  

Another  important insight from the weak coupling approach concerns the form of the gap function.  All such calculations~\cite{raghuandme,nomura,Ronny,Kallin,PhysRevLett.105.136401,weejee} indicate a tendency toward pair-wave functions with substantially more complex structure than the minimal functions we considered above.  In fact, deep gap minima not required by symmetry arise ubiquitously in such studies.  %\ronny
{ %\color{green}
 Such ``accidental near-nodes'' were invoked in earlier attempts to reconcile $p_x+i p_y$ and d+is  SC states with experiments that suggest the existence of line  nodes. % Previously,
The possibility of $d_{x^2-y^2}+is$ %most prominently came up
pairing near a point of accidental degeneracy of the two ordering tendencies has been considered previously  in the context of iron pnictides~\cite{PhysRevB.85.180502}, %claimed to arise from an accidentally nodal extended s-wave and a $d$-wave state~\cite{PhysRevB.85.180502}, 
 %and has also recently surfaced for the ruthenates
 and more recently in the context of SRO~\cite{PhysRevLett.123.247001}. While in our present scenario, there are symmetry protected true line-nodes even in the $d_{x^2-y^2}+i g_{xy(x^2-y^2)}$ SC phase that would be absent in its  $d_{x^2-y^2}+i s$-wave cousin, it is possible that further structure of the gaps could be invoked to produce near-nodes at the requisite locations.  %, %and possibly the presence of additional near-nodes, can play a role in determining a host of physical properties. 
  (We also plan to address this in future studies~\cite{inprep}.)} 
 
 \section{%Experimental observations we have neglected
 Experimental issues that require further study}
  \label{wrong}

One of the strongest pieces of evidence for TRSB is the onset of the Kerr effect. On symmetry grounds, the Kerr effect vanishes in any state that does not break time-reversal symmetry {\rm and} all vertical mirror plane symmetries.  While a $d_{x^2-y^2}+i g_{xy(x^2-y^2)}$ SC breaks time-reversal symmetry, it preserves mirror symmetry through the planes perpendicular to $(1,1,0)$ and $(1, -1,0)$.  Since the Kerr measurements are performed in an experimental geometry that likewise preserves these mirror symmetries, the proposed state, by itself, cannot account for the observed Kerr signal.  Disorder, especially the sort of strain disorder we have already invoked, generically breaks these mirror symmetries (but not time-reversal symmetry).  Consequently, we conjecture that the observed Kerr measurements are likely consistent with the present scenario, but this is an issue that deserves further analysis.
%\ronny{On a related note, 

A variety of other experimental observations need to be addressed.  %While not repeated by other experimental groups, %rudimentary
Evidence of  triplet pairing has been adduced from experiments on  214/113 strontium ruthenate heterostructure tunneling~\cite{anwar2019observation}, of odd-parity pairing from phase sensitive Josephson tunneling~\cite{josephson}, of horizontal line nodes from field dependent specific heat measurements~\cite{kittaka2018searching}, and of half quantum vortices from cantilever magnetometry~\cite{jang2011observation}.  All these observatioons pose some challenge to be reconciled with our ansatz, and thus  deserve further analysis and revisiting from both an experimental and  a theoretical viewpoint. 

%\ronny{summarize those three experiments in two or three sentences}
 
% Maeno work on tunnelling between SRO 214 and SRO 113 provides evidence of triplet pairing. \cite{anwar2019observation}
 
% Ying Liu phase sensitive tunneling experiment that shows evidence of odd parity pairing \cite{josephson}
 
% I am not sure what to say about Rafi Budakian's experiment \cite{jang2011observation}
 
% The  field  dependent specific heat measurements that claimed  evidence for horizontal line nodes \cite{kittaka2018searching}.

 \section{Proposals for future experiments}
 \label{future}
 
 Because the proposed degeneracy between d and g is accidental, it is not plausible that it would turn out to be exact under conditions of zero  strain and zero disorder.  Thus, we would expect there to be a small but non-zero splitting between $T_c$ and $T_{trsb}$ - however, if this is small, it  might be difficult to detect under circumstances in which the root mean square variance of $h_0$  or $h_1$ ($\sigma_0$ or $\sigma_1$) produce variations of $T_c$ and $T_{trsb}$ that are larger  than the intrinsic splitting. %, $|\alpha_1|$. 
  A possibly more  directly testable aspect of  this proposal, however, is  that the application of any strain - regardless of its symmetry - would be  expected to enhance this splitting.  Specifically,  in the absence of disorder, and under the assumption that %$\alpha_1=0$, 
  the degeneracy is effective exact at zero strain, we would expect a strain-induced  splitting 
   \bea
 &&T_c - T_{trsb}  = \sqrt{\delta T_1^2+\delta T_2^2} \\
 &&\delta T_1 \sim \left |\Gamma_1^s\epsilon_s +  \Gamma_1^{s,s}\epsilon_s^2\right| + \Gamma_1^{xy,xy}\epsilon_{xy}^2 + \Gamma_1^{x^2-y^2,x^2-y^2}\epsilon_{x^2-y^2}^2 \nonumber \\
&& \delta T_2 \sim \left |\Gamma_2^{xy}\epsilon_{xy}+\Gamma_2^{xy,s} \epsilon_{xy}\epsilon_s\right|
 \eea
 In particular, the prediction that a symmetry preserving strain, $\epsilon_s$, will lead  to  a splitting of the transition directly reflects the  assumption that it is an accidental degeneracy - rather than a symmetry protected one - that  is responsible  for  the  two-component character  of the order.
 
 Similarly, disorder (especially short-range correlated disorder) is expected to have a substantial impact on $T_c$ for any unconventional SC.  In the present case, because they are not symmetry related, it should have a different effect on the d and g wave components, thus leading to a split transition.  Since the g-wave order has more (symmetry related) nodes on the Fermi surface, one would probably expect disorder to suppress the g-wave component more strongly, thus leading to a case in which there is pure $d$-wave pairing below $T_c$ and a lower temperature transition to a $d_{x^2-y^2}+i g_{xy(x^2-y^2)}$ SC at $T_{trsb} < T_c$.
 
 \acknowledgements
 We would like to acknowledge helpful discussions with E.~Berg, S.~Brown, S. Raghu, T.~Schwemmer,  G.~Tarjus and especially with A.~P.~Mackenzie. S.~A.~K. and R.~T. acknowledge support by the Humboldt Foundation. R.~T. thanks Clara Theodora for inspiration. The work in W\"urzburg is funded by the Deutsche Forschungsgemeinschaft (DFG, German Research Foundation) through project-id 258499086 - SFB 1170 and through the W\"urzburg-Dresden Cluster of Excellence on Complexity and Topology in Quantum Matter –ct.qmat project-id 39085490 - EXC 2147. B.J.R. acknowledges funding from the National Science Foundation under Grant No. DMR-1752784.  SAK was supported in part by NSF grant \# DMR- 1608055 at Stanford. A.C.Y. acknowledges support by the Department of Energy, Office of Science, Basic Energy Sciences, Materials Sciences and Engineering Division, under Contract DE-AC02-76SF00515..

%\bibliographystyle{unsrtnat}
%\bibliography{biblio}

\begin{thebibliography}{50}
\providecommand{\natexlab}[1]{#1}
\providecommand{\url}[1]{\texttt{#1}}
\expandafter\ifx\csname urlstyle\endcsname\relax
  \providecommand{\doi}[1]{doi: #1}\else
  \providecommand{\doi}{doi: \begingroup \urlstyle{rm}\Url}\fi

\bibitem[Maeno et~al.(1994)Maeno, Hashimoto, Yoshida, Nishizaki, Fujita,
  Bednorz, and Lichtenberg]{firstpaper}
Y.~Maeno, H.~Hashimoto, K.~Yoshida, S.~Nishizaki, T.~Fujita, J.~G. Bednorz, and
  F.~Lichtenberg.
\newblock Superconductivity in a layered perovskite without copper.
\newblock \emph{Nature}, 372\penalty0 (6506):\penalty0 532--534, 1994.

\bibitem[Ishida et~al.(1997)Ishida, Kitaoka, Asayama, Ikeda, Nishizaki, Maeno,
  Yoshida, and Fujita]{ishida1997anisotropic}
K.~Ishida, Y.~Kitaoka, K.~Asayama, S.~Ikeda, S.~Nishizaki, Y.~Maeno,
  K.~Yoshida, and T.~Fujita.
\newblock Anisotropic pairing in superconducting
  ${\mathrm{sr}}_{2}{\mathrm{ruo}}_{4}:$ ru nmr and nqr studies.
\newblock \emph{Phys. Rev. B}, 56:\penalty0 R505--R508, Jul 1997.
\newblock \doi{10.1103/PhysRevB.56.R505}.
\newblock URL \url{https://link.aps.org/doi/10.1103/PhysRevB.56.R505}.

\bibitem[Mackenzie et~al.(1998)Mackenzie, Haselwimmer, Tyler, Lonzarich, Mori,
  Nishizaki, and Maeno]{mackenzieunconventional}
AP~Mackenzie, RKW Haselwimmer, AW~Tyler, GG~Lonzarich, Y~Mori, S~Nishizaki, and
  Y~Maeno.
\newblock Extremely strong dependence of superconductivity on disorder in sr 2
  ruo 4.
\newblock \emph{Physical review letters}, 80\penalty0 (1):\penalty0 161, 1998.

\bibitem[Mackenzie et~al.({2017})Mackenzie, Scaffidi, Hicks, and
  Maeno]{sroreview}
Andrew~P. Mackenzie, Thomas Scaffidi, Clifford~W. Hicks, and Yoshiteru Maeno.
\newblock {Even odder after twenty-three years: the superconducting order
  parameter puzzle of Sr2RuO4}.
\newblock \emph{{NPJ Quantum Materials}}, {2}, {JUL 18} {2017}.
\newblock ISSN {2397-4648}.

\bibitem[Rice and Sigrist(1995)]{Rice}
T~M Rice and M~Sigrist.
\newblock Sr2ruo4: an electronic analogue of3he?
\newblock \emph{Journal of Physics: Condensed Matter}, 7\penalty0
  (47):\penalty0 L643--L648, nov 1995.
\newblock \doi{10.1088/0953-8984/7/47/002}.
\newblock URL \url{https://doi.org/10.1088%2F0953-8984%2F7%2F47%2F002}.

\bibitem[Baskaran(1996)]{Baskaran}
G.~Baskaran.
\newblock Why is sr$_2$ruo$_4$ not a high t$_c$ superconductor? electron
  correlation, hund's coupling and p-wave superconductivity.
\newblock \emph{Physica B: Condensed Matter}, 223:\penalty0 490, 1996.

\bibitem[Luke et~al.(1998)Luke, Fudamoto, Kojima, Larkin, Merrin, Nachumi,
  Uemura, Maeno, Mao, Mori, et~al.]{musr}
G~Ml Luke, Y~Fudamoto, KM~Kojima, MI~Larkin, J~Merrin, B~Nachumi, YJ~Uemura,
  Y~Maeno, ZQ~Mao, Y~Mori, et~al.
\newblock Time-reversal symmetry-breaking superconductivity in sr 2 ruo 4.
\newblock \emph{Nature}, 394\penalty0 (6693):\penalty0 558, 1998.

\bibitem[Xia et~al.(2006)Xia, Maeno, Beyersdorf, Fejer, and Kapitulnik]{kerr}
Jing Xia, Yoshiteru Maeno, Peter~T Beyersdorf, MM~Fejer, and Aharon Kapitulnik.
\newblock High resolution polar kerr effect measurements of sr 2 ruo 4:
  Evidence for broken time-reversal symmetry in the superconducting state.
\newblock \emph{Physical review letters}, 97\penalty0 (16):\penalty0 167002,
  2006.

\bibitem[Kidwingira et~al.(2006)Kidwingira, Strand, Van~Harlingen, and
  Maeno]{Kidwingira1267}
Francoise Kidwingira, J.~D. Strand, D.~J. Van~Harlingen, and Yoshiteru Maeno.
\newblock Dynamical superconducting order parameter domains in sr2ruo4.
\newblock \emph{Science}, 314\penalty0 (5803):\penalty0 1267--1271, 2006.
\newblock ISSN 0036-8075.
\newblock \doi{10.1126/science.1133239}.
\newblock URL \url{https://science.sciencemag.org/content/314/5803/1267}.

\bibitem[NishiZaki et~al.(2000)NishiZaki, Maeno, and
  Mao]{doi:10.1143/JPSJ.69.572}
Shuji NishiZaki, Yoshiteru Maeno, and Zhiqiang Mao.
\newblock Changes in the superconducting state of sr 2ruo 4 under magnetic
  fields probed by specific heat.
\newblock \emph{Journal of the Physical Society of Japan}, 69\penalty0
  (2):\penalty0 572--578, 2000.
\newblock \doi{10.1143/JPSJ.69.572}.
\newblock URL \url{https://doi.org/10.1143/JPSJ.69.572}.

\bibitem[Bonalde et~al.(2000)Bonalde, Yanoff, Salamon, Van~Harlingen, Chia,
  Mao, and Maeno]{penetrationdepth}
I.~Bonalde, Brian~D. Yanoff, M.~B. Salamon, D.~J. Van~Harlingen, E.~M.~E. Chia,
  Z.~Q. Mao, and Y.~Maeno.
\newblock Temperature dependence of the penetration depth in
  ${\mathrm{sr}}_{2}{\mathrm{ruo}}_{4}$: Evidence for nodes in the gap
  function.
\newblock \emph{Phys. Rev. Lett.}, 85:\penalty0 4775--4778, Nov 2000.
\newblock \doi{10.1103/PhysRevLett.85.4775}.
\newblock URL \url{https://link.aps.org/doi/10.1103/PhysRevLett.85.4775}.

\bibitem[Lupien et~al.(2001)Lupien, MacFarlane, Proust, Taillefer, Mao, and
  Maeno]{lupien2001ultrasound}
C.~Lupien, W.~A. MacFarlane, Cyril Proust, Louis Taillefer, Z.~Q. Mao, and
  Y.~Maeno.
\newblock Ultrasound attenuation in ${\mathrm{sr}}_{2}{\mathrm{ruo}}_{4}$: An
  angle-resolved study of the superconducting gap function.
\newblock \emph{Phys. Rev. Lett.}, 86:\penalty0 5986--5989, Jun 2001.
\newblock \doi{10.1103/PhysRevLett.86.5986}.
\newblock URL \url{https://link.aps.org/doi/10.1103/PhysRevLett.86.5986}.

\bibitem[Hassinger et~al.(2017)Hassinger, Bourgeois-Hope, Taniguchi, de~Cotret,
  Grissonnanche, Anwar, Maeno, Doiron-Leyraud, and
  Taillefer]{thermalconductivity}
Elena Hassinger, Patrick Bourgeois-Hope, Haruka Taniguchi, S~Ren{\'e}
  de~Cotret, Gael Grissonnanche, M~Shahbaz Anwar, Yoshiteru Maeno, Nicolas
  Doiron-Leyraud, and Louis Taillefer.
\newblock Vertical line nodes in the superconducting gap structure of sr 2 ruo
  4.
\newblock \emph{Physical Review X}, 7\penalty0 (1):\penalty0 011032, 2017.

\bibitem[Firmo et~al.(2013)Firmo, Lederer, Lupien, Mackenzie, Davis, and
  Kivelson]{Irmo}
I.~A. Firmo, S.~Lederer, C.~Lupien, A.~P. Mackenzie, J.~C. Davis, and S.~A.
  Kivelson.
\newblock Evidence from tunneling spectroscopy for a quasi-one-dimensional
  origin of superconductivity in sr${}_{2}$ruo${}_{4}$.
\newblock \emph{Phys. Rev. B}, 88:\penalty0 134521, Oct 2013.
\newblock \doi{10.1103/PhysRevB.88.134521}.
\newblock URL \url{https://link.aps.org/doi/10.1103/PhysRevB.88.134521}.

\bibitem[Sharma et~al.(2019)Sharma, Edkins, Wang, Kostin, Sow, Maeno,
  Mackenzie, Davis, and Madhavan]{stm2}
Rahul Sharma, Stephen~D Edkins, Zhenyu Wang, Andrey Kostin, Chanchal Sow,
  Yoshiteru Maeno, Andrew~P Mackenzie, JC~Davis, and Vidya Madhavan.
\newblock Momentum resolved superconducting energy gaps of sr $ \_2 $ ruo $ \_4
  $ from quasiparticle interference imaging.
\newblock \emph{arXiv preprint arXiv:1912.02798}, 2019.

\bibitem[Lupien(2002)]{lupien2002ultrasound}
Christian Lupien.
\newblock \emph{Ultrasound attenuation in the unconventional superconductor
  Sr2RuO4}.
\newblock PhD Thesis, 2002.

\bibitem[Ghosh et~al.(2020)Ghosh, Shekhter, Jerzembeck, Kikugawa, Sokolov,
  Mackenzie, Hicks, and Ramshaw]{bradspaper}
Sayak Ghosh, Arkady Shekhter, F.~Jerzembeck, N.~Kikugawa, Dimitry~A. Sokolov,
  A.~P. Mackenzie, Clifford~W. Hicks, and B.~J. Ramshaw.
\newblock Thermodynamic evidence for a two-component superconducting order
  parameter in sr2ruo4.
\newblock \emph{Manuscript in preparation}, 2020.

\bibitem[Hicks et~al.(2014)Hicks, Brodsky, Yelland, Gibbs, Bruin, Barber,
  Edkins, Nishimura, Yonezawa, Maeno, et~al.]{hicksstrain}
Clifford~W Hicks, Daniel~O Brodsky, Edward~A Yelland, Alexandra~S Gibbs, Jan~AN
  Bruin, Mark~E Barber, Stephen~D Edkins, Keigo Nishimura, Shingo Yonezawa,
  Yoshiteru Maeno, et~al.
\newblock Strong increase of tc of sr2ruo4 under both tensile and compressive
  strain.
\newblock \emph{Science}, 344\penalty0 (6181):\penalty0 283--285, 2014.

\bibitem[Deguchi et~al.(2002{\natexlab{a}})Deguchi, A.~Tanatar, Mao, Ishiguro,
  and Maeno]{doi:10.1143/JPSJ.71.2839}
Kazuhiko Deguchi, Makariy A.~Tanatar, Zhiqiang Mao, Takehiko Ishiguro, and
  Yoshiteru Maeno.
\newblock Superconducting double transition and the upper critical field limit
  of sr2ruo4 in parallel magnetic fields.
\newblock \emph{Journal of the Physical Society of Japan}, 71\penalty0
  (12):\penalty0 2839--2842, 2002{\natexlab{a}}.
\newblock \doi{10.1143/JPSJ.71.2839}.
\newblock URL \url{https://doi.org/10.1143/JPSJ.71.2839}.

\bibitem[Nelson et~al.(2004)Nelson, Mao, Maeno, and Liu]{josephson}
KD~Nelson, ZQ~Mao, Y~Maeno, and Ying Liu.
\newblock Odd-parity superconductivity in sr2ruo4.
\newblock \emph{Science}, 306\penalty0 (5699):\penalty0 1151--1154, 2004.

\bibitem[Pustogow et~al.(2019{\natexlab{a}})Pustogow, Luo, Chronister, Su,
  Sokolov, Jerzembeck, Mackenzie, Hicks, Kikugawa, Raghu, et~al.]{brown}
A~Pustogow, Yongkang Luo, A~Chronister, Y-S Su, DA~Sokolov, F~Jerzembeck,
  AP~Mackenzie, CW~Hicks, N~Kikugawa, S~Raghu, et~al.
\newblock Constraints on the superconducting order parameter in sr 2 ruo 4 from
  oxygen-17 nuclear magnetic resonance.
\newblock \emph{Nature}, 574\penalty0 (7776):\penalty0 72--75,
  2019{\natexlab{a}}.

\bibitem[Ishida et~al.(2019)Ishida, Manago, and Maeno]{newjapan}
Kenji Ishida, Masahiro Manago, and Yoshiteru Maeno.
\newblock Reduction of the $^{17}$ o knight shift in the superconducting state
  and the heat-up effect by nmr pulses on sr $ \_2 $ ruo $ \_4$.
\newblock \emph{arXiv preprint arXiv:1907.12236}, 2019.

\bibitem[Deguchi et~al.(2002{\natexlab{b}})Deguchi, Tanatar, Mao, Ishiguro, and
  Maeno]{paulilimit}
Kazuhiko Deguchi, Makariy~A Tanatar, Zhiqiang Mao, Takehiko Ishiguro, and
  Yoshiteru Maeno.
\newblock Superconducting double transition and the upper critical field limit
  of sr 2 ruo 4 in parallel magnetic fields.
\newblock \emph{Journal of the Physical Society of Japan}, 71\penalty0
  (12):\penalty0 2839--2842, 2002{\natexlab{b}}.

\bibitem[Watson et~al.(2018)Watson, Gibbs, Mackenzie, Hicks, and Moler]{kam}
Christopher~A. Watson, Alexandra~S. Gibbs, Andrew~P. Mackenzie, Clifford~W.
  Hicks, and Kathryn~A. Moler.
\newblock Micron-scale measurements of low anisotropic strain response of local
  ${T}_{c}$ in ${\mathrm{sr}}_{2}{\mathrm{ruo}}_{4}$.
\newblock \emph{Phys. Rev. B}, 98:\penalty0 094521, Sep 2018.
\newblock \doi{10.1103/PhysRevB.98.094521}.
\newblock URL \url{https://link.aps.org/doi/10.1103/PhysRevB.98.094521}.

\bibitem[Kirtley et~al.(2007)Kirtley, Kallin, Hicks, Kim, Liu, Moler, Maeno,
  and Nelson]{KAMattheedge}
J.~R. Kirtley, C.~Kallin, C.~W. Hicks, E.-A. Kim, Y.~Liu, K.~A. Moler,
  Y.~Maeno, and K.~D. Nelson.
\newblock Upper limit on spontaneous supercurrents in
  ${\mathrm{sr}}_{2}\mathrm{Ru}{\mathrm{o}}_{4}$.
\newblock \emph{Phys. Rev. B}, 76:\penalty0 014526, Jul 2007.
\newblock \doi{10.1103/PhysRevB.76.014526}.
\newblock URL \url{https://link.aps.org/doi/10.1103/PhysRevB.76.014526}.

\bibitem[Ghosh et~al.(2019)Ghosh, Matty, Baumbach, Bauer, Modic, Shekhter,
  Mydosh, Kim, and Ramshaw]{ghosh2019single}
Sayak Ghosh, Michael Matty, Ryan Baumbach, Eric~D Bauer, KA~Modic, Arkady
  Shekhter, JA~Mydosh, Eun-Ah Kim, and BJ~Ramshaw.
\newblock Single-component order parameter in uru $ \_2 $ si $ \_2 $ uncovered
  by resonant ultrasound spectroscopy and machine learning.
\newblock \emph{arXiv preprint arXiv:1903.00552}, 2019.

\bibitem[{Grinenko} et~al.(2020){Grinenko}, {Ghosh}, {Sarkar}, {Orain},
  {Nikitin}, {Elender}, {Das}, {Guguchia}, {Br{\"u}ckner}, {Barber}, {Park},
  {Kikugawa}, {Sokolov}, {Bobowski}, {Miyoshi}, {Maeno}, {Mackenzie},
  {Luetkens}, {Hicks}, and {Klauss}]{newmusr}
Vadim {Grinenko}, Shreenanda {Ghosh}, Rajib {Sarkar}, Jean-Christophe {Orain},
  Artem {Nikitin}, Matthias {Elender}, Debarchan {Das}, Zurab {Guguchia}, Felix
  {Br{\"u}ckner}, Mark~E. {Barber}, Joonbum {Park}, Naoki {Kikugawa}, Dmitry~A.
  {Sokolov}, Jake~S. {Bobowski}, Takuto {Miyoshi}, Yoshiteru {Maeno}, Andrew~P.
  {Mackenzie}, Hubertus {Luetkens}, Clifford~W. {Hicks}, and Hans-Henning
  {Klauss}.
\newblock {Split superconducting and time-reversal symmetry-breaking
  transitions, and magnetic order in Sr$_2$RuO$_4$ under uniaxial stress}.
\newblock \emph{arXiv e-prints}, art. arXiv:2001.08152, Jan 2020.

\bibitem[Li et~al.(2019)Li, Kikugawa, Sokolov, Jerzembeck, Gibbs, Maeno, Hicks,
  Nicklas, and Mackenzie]{li2019high}
Y-S Li, N~Kikugawa, DA~Sokolov, F~Jerzembeck, AS~Gibbs, Y~Maeno, CW~Hicks,
  M~Nicklas, and AP~Mackenzie.
\newblock High precision heat capacity measurements on sr2ruo4 under uniaxial
  pressure.
\newblock \emph{arXiv preprint arXiv:1906.07597}, 2019.

\bibitem[Pustogow et~al.(2019{\natexlab{b}})Pustogow, Luo, Chronister, Su,
  Sokolov, Jerzembeck, Mackenzie, Hicks, Kikugawa, Raghu, Bauer, and
  Brown]{stuartspaper}
A.~Pustogow, Yongkang Luo, A.~Chronister, Y.~S. Su, D.~A. Sokolov,
  F.~Jerzembeck, A.~P. Mackenzie, C.~W. Hicks, N.~Kikugawa, S.~Raghu, E.~D.
  Bauer, and S.~E. Brown.
\newblock Constraints on the superconducting order parameter in sr2ruo4 from
  oxygen-17 nuclear magnetic resonance.
\newblock \emph{Nature}, 574\penalty0 (7776):\penalty0 72--75,
  2019{\natexlab{b}}.

\bibitem[\ifmmode \check{Z}\else \v{Z}\fi{}uti\ifmmode~\acute{c}\else
  \'{c}\fi{} and Mazin(2005)]{PhysRevLett.95.217004}
Igor \ifmmode \check{Z}\else \v{Z}\fi{}uti\ifmmode~\acute{c}\else \'{c}\fi{}
  and Igor Mazin.
\newblock Phase-sensitive tests of the pairing state symmetry in
  ${\mathrm{sr}}_{2}{\mathrm{ruo}}_{4}$.
\newblock \emph{Phys. Rev. Lett.}, 95:\penalty0 217004, Nov 2005.
\newblock \doi{10.1103/PhysRevLett.95.217004}.
\newblock URL \url{https://link.aps.org/doi/10.1103/PhysRevLett.95.217004}.

\bibitem[Puetter and Kee(2012)]{Puetter_2012}
Christoph~M. Puetter and Hae-Young Kee.
\newblock Identifying spin-triplet pairing in spin-orbit coupled multi-band
  superconductors.
\newblock \emph{{EPL} (Europhysics Letters)}, 98\penalty0 (2):\penalty0 27010,
  apr 2012.
\newblock \doi{10.1209/0295-5075/98/27010}.
\newblock URL \url{https://doi.org/10.1209%2F0295-5075%2F98%2F27010}.

\bibitem[Suh et~al.()Suh, Menke, Brydon, Timm, Ramirez, and Agterberg]{Suh}
H.~G. Suh, H.~Menke, P.~M.~R. Brydon, C.~Timm, A.~Ramirez, and D.~F. Agterberg.
\newblock Stabilizing even-parity chiral superconductivity in sr$_2$ruo$_4$.
\newblock arXiv:1912.09525.

\bibitem[Yu and Raghu(2019)]{yue}
Yue Yu and S.~Raghu.
\newblock Effect of strain inhomogeneity on a chiral $p$-wave superconductor.
\newblock \emph{Phys. Rev. B}, 100:\penalty0 094517, Sep 2019.
\newblock \doi{10.1103/PhysRevB.100.094517}.
\newblock URL \url{https://link.aps.org/doi/10.1103/PhysRevB.100.094517}.

\bibitem[Raghu et~al.(2010{\natexlab{a}})Raghu, Kapitulnik, and
  Kivelson]{raghuandme}
S.~Raghu, A.~Kapitulnik, and S.~A. Kivelson.
\newblock Hidden quasi-one-dimensional superconductivity in
  ${\mathrm{sr}}_{2}{\mathrm{ruo}}_{4}$.
\newblock \emph{Phys. Rev. Lett.}, 105:\penalty0 136401, Sep
  2010{\natexlab{a}}.
\newblock \doi{10.1103/PhysRevLett.105.136401}.
\newblock URL \url{https://link.aps.org/doi/10.1103/PhysRevLett.105.136401}.

\bibitem[Raghu et~al.(2010{\natexlab{b}})Raghu, Kapitulnik, and
  Kivelson]{PhysRevLett.105.136401}
S.~Raghu, A.~Kapitulnik, and S.~A. Kivelson.
\newblock Hidden quasi-one-dimensional superconductivity in
  ${\mathrm{sr}}_{2}{\mathrm{ruo}}_{4}$.
\newblock \emph{Phys. Rev. Lett.}, 105:\penalty0 136401, Sep
  2010{\natexlab{b}}.
\newblock \doi{10.1103/PhysRevLett.105.136401}.
\newblock URL \url{https://link.aps.org/doi/10.1103/PhysRevLett.105.136401}.

\bibitem[Steppke et~al.(2017)Steppke, Zhao, Barber, Scaffidi, Jerzembeck,
  Rosner, Gibbs, Maeno, Simon, Mackenzie, and Hicks]{Steppkeeaaf9398}
Alexander Steppke, Lishan Zhao, Mark~E. Barber, Thomas Scaffidi, Fabian
  Jerzembeck, Helge Rosner, Alexandra~S. Gibbs, Yoshiteru Maeno, Steven~H.
  Simon, Andrew~P. Mackenzie, and Clifford~W. Hicks.
\newblock Strong peak in tc of sr2ruo4 under uniaxial pressure.
\newblock \emph{Science}, 355\penalty0 (6321), 2017.
\newblock ISSN 0036-8075.
\newblock \doi{10.1126/science.aaf9398}.
\newblock URL \url{https://science.sciencemag.org/content/355/6321/eaaf9398}.

\bibitem[Zhang et~al.(2018)Zhang, Huang, Yang, and Yao]{hong}
Li-Da Zhang, Wen Huang, Fan Yang, and Hong Yao.
\newblock Superconducting pairing in ${\mathrm{sr}}_{2}{\mathrm{ruo}}_{4}$ from
  weak to intermediate coupling.
\newblock \emph{Phys. Rev. B}, 97:\penalty0 060510, Feb 2018.
\newblock \doi{10.1103/PhysRevB.97.060510}.
\newblock URL \url{https://link.aps.org/doi/10.1103/PhysRevB.97.060510}.

\bibitem[{Wang, Q. H.} et~al.(2013){Wang, Q. H.}, {Platt, C.}, {Yang, Y.},
  {Honerkamp, C.}, {Zhang, F. C.}, {Hanke, W.}, {Rice, T. M.}, and {Thomale,
  R.}]{Ronny}
{Wang, Q. H.}, {Platt, C.}, {Yang, Y.}, {Honerkamp, C.}, {Zhang, F. C.},
  {Hanke, W.}, {Rice, T. M.}, and {Thomale, R.}
\newblock Theory of superconductivity in a three-orbital model of sr2ruo4.
\newblock \emph{EPL}, 104\penalty0 (1):\penalty0 17013, 2013.
\newblock \doi{10.1209/0295-5075/104/17013}.
\newblock URL \url{https://doi.org/10.1209/0295-5075/104/17013}.

\bibitem[Wang et~al.(2019)Wang, Zhang, Zhang, and Wang]{PhysRevLett.122.027002}
Wan-Sheng Wang, Cong-Cong Zhang, Fu-Chun Zhang, and Qiang-Hua Wang.
\newblock Theory of chiral $p$-wave superconductivity with near nodes for
  ${\mathrm{sr}}_{2}{\mathrm{ruo}}_{4}$.
\newblock \emph{Phys. Rev. Lett.}, 122:\penalty0 027002, Jan 2019.
\newblock \doi{10.1103/PhysRevLett.122.027002}.
\newblock URL \url{https://link.aps.org/doi/10.1103/PhysRevLett.122.027002}.

\bibitem[Raghu et~al.(2012)Raghu, Berg, Chubukov, and
  Kivelson]{PhysRevB.85.024516}
S.~Raghu, E.~Berg, A.~V. Chubukov, and S.~A. Kivelson.
\newblock Effects of longer-range interactions on unconventional
  superconductivity.
\newblock \emph{Phys. Rev. B}, 85:\penalty0 024516, Jan 2012.
\newblock \doi{10.1103/PhysRevB.85.024516}.
\newblock URL \url{https://link.aps.org/doi/10.1103/PhysRevB.85.024516}.

\bibitem[inp()]{inprep}
in preparation.

\bibitem[Schnell et~al.(2006)Schnell, Mazin, and Liu]{PhysRevB.74.184503}
I.~Schnell, I.~I. Mazin, and Amy~Y. Liu.
\newblock Unconventional superconducting pairing symmetry induced by phonons.
\newblock \emph{Phys. Rev. B}, 74:\penalty0 184503, Nov 2006.
\newblock \doi{10.1103/PhysRevB.74.184503}.
\newblock URL \url{https://link.aps.org/doi/10.1103/PhysRevB.74.184503}.

\bibitem[Nomura and Yamada(2002)]{nomura}
Takuji Nomura and Kosaku Yamada.
\newblock Roles of electron correlations in the spin-triplet superconductivity
  of sr2ruo4.
\newblock \emph{Journal of the Physical Society of Japan}, 71\penalty0
  (8):\penalty0 1993--2004, 2002.
\newblock \doi{10.1143/JPSJ.71.1993}.
\newblock URL \url{https://doi.org/10.1143/JPSJ.71.1993}.

\bibitem[Kallin and Berlinsky(2009)]{Kallin}
C~Kallin and A~J Berlinsky.
\newblock Is sr2ruo4a chiral p-wave superconductor?
\newblock \emph{Journal of Physics: Condensed Matter}, 21\penalty0
  (16):\penalty0 164210, mar 2009.
\newblock \doi{10.1088/0953-8984/21/16/164210}.
\newblock URL \url{https://doi.org/10.1088%2F0953-8984%2F21%2F16%2F164210}.

\bibitem[Cho et~al.(2013)Cho, Thomale, Raghu, and Kivelson]{weejee}
Weejee Cho, Ronny Thomale, Srinivas Raghu, and Steven~A. Kivelson.
\newblock Band structure effects on the superconductivity in hubbard models.
\newblock \emph{Phys. Rev. B}, 88:\penalty0 064505, Aug 2013.
\newblock \doi{10.1103/PhysRevB.88.064505}.
\newblock URL \url{https://link.aps.org/doi/10.1103/PhysRevB.88.064505}.

\bibitem[Platt et~al.(2012)Platt, Thomale, Honerkamp, Zhang, and
  Hanke]{PhysRevB.85.180502}
Christian Platt, Ronny Thomale, Carsten Honerkamp, Shou-Cheng Zhang, and Werner
  Hanke.
\newblock Mechanism for a pairing state with time-reversal symmetry breaking in
  iron-based superconductors.
\newblock \emph{Phys. Rev. B}, 85:\penalty0 180502, May 2012.
\newblock \doi{10.1103/PhysRevB.85.180502}.
\newblock URL \url{https://link.aps.org/doi/10.1103/PhysRevB.85.180502}.

\bibitem[R\o{}mer et~al.(2019)R\o{}mer, Scherer, Eremin, Hirschfeld, and
  Andersen]{PhysRevLett.123.247001}
A.~T. R\o{}mer, D.~D. Scherer, I.~M. Eremin, P.~J. Hirschfeld, and B.~M.
  Andersen.
\newblock Knight shift and leading superconducting instability from spin
  fluctuations in ${\mathrm{sr}}_{2}{\mathrm{ruo}}_{4}$.
\newblock \emph{Phys. Rev. Lett.}, 123:\penalty0 247001, Dec 2019.
\newblock \doi{10.1103/PhysRevLett.123.247001}.
\newblock URL \url{https://link.aps.org/doi/10.1103/PhysRevLett.123.247001}.

\bibitem[Anwar et~al.(2019)Anwar, Kunieda, Ishiguro, Lee, Olthof, Robinson,
  Yonezawa, Noh, and Maeno]{anwar2019observation}
M.~S. Anwar, M.~Kunieda, R.~Ishiguro, S.~R. Lee, L.~A. B.~Olde Olthof, J.~W.~A.
  Robinson, S.~Yonezawa, T.~W. Noh, and Y.~Maeno.
\newblock Observation of superconducting gap spectra of long-range proximity
  effect in
  $\mathrm{Au}/{\mathrm{srtio}}_{3}/{\mathrm{srruo}}_{3}/{\mathrm{sr}}_{2}{\mathrm{ruo}}_{4}$
  tunnel junctions.
\newblock \emph{Phys. Rev. B}, 100:\penalty0 024516, Jul 2019.
\newblock \doi{10.1103/PhysRevB.100.024516}.
\newblock URL \url{https://link.aps.org/doi/10.1103/PhysRevB.100.024516}.

\bibitem[Kittaka et~al.(2018)Kittaka, Nakamura, Sakakibara, Kikugawa,
  Terashima, Uji, Sokolov, Mackenzie, Irie, Tsutsumi,
  et~al.]{kittaka2018searching}
Shunichiro Kittaka, Shota Nakamura, Toshiro Sakakibara, Naoki Kikugawa, Taichi
  Terashima, Shinya Uji, Dmitry~A Sokolov, Andrew~P Mackenzie, Koki Irie,
  Yasumasa Tsutsumi, et~al.
\newblock Searching for gap zeros in sr2ruo4 via field-angle-dependent
  specific-heat measurement.
\newblock \emph{Journal of the Physical Society of Japan}, 87\penalty0
  (9):\penalty0 093703, 2018.

\bibitem[Jang et~al.(2011)Jang, Ferguson, Vakaryuk, Budakian, Chung, Goldbart,
  and Maeno]{jang2011observation}
J~Jang, DG~Ferguson, V~Vakaryuk, Raffi Budakian, SB~Chung, PM~Goldbart, and
  Y~Maeno.
\newblock Observation of half-height magnetization steps in sr2ruo4.
\newblock \emph{Science}, 331\penalty0 (6014):\penalty0 186--188, 2011.

\end{thebibliography}

%\begin{thebibliography}{12}
%\expandafter\ifx\csname natexlab\endcsname\relax\def\natexlab#1{#1}\fi
%\expandafter\ifx\csname bibnamefont\endcsname\relax
  %\def\bibnamefont#1{#1}\fi
%\expandafter\ifx\csname bibfnamefont\endcsname\relax
  %\def\bibfnamefont#1{#1}\fi
%\expandafter\ifx\csname citenamefont\endcsname\relax
  %\def\citenamefont#1{#1}\fi
%\expandafter\ifx\csname url\endcsname\relax
  %\def\url#1{\texttt{#1}}\fi
%\expandafter\ifx\csname urlprefix\endcsname\relax\def\urlprefix{URL }\fi
%\providecommand{\bibinfo}[2]{#2}
%\providecommand{\eprint}[2][]{\url{#2}}
%
%\bibitem{sroreview} For a review, see Mackenzie review in npj QM
%
%\bibitem{musr}  muSR evidence
%\bibitem{kerr} Kerr  evidence
%\bibitem{josephson} Josephson evidence (van Harlingen)
%\bibitem{specificheat}  Specific heat evidence of nodes
%\bibitem{thermalconductivity} Thermal conductivity (Louis)
%\bibitem{stm1} Firmo et al STM
%\bibitem{stm2} New Davis STM papper
%\bibitem{bradspaper} Brad's paper
%\bibitem{hicksstrain} Hick's Strain 
%\bibitem{kamstrain} Kams Strain paper
%\bibitem{inplanefielddoesntsplittransition} No splitting of transition by in-plane field.
%\bibitem{brown} Stuart's paper
%\bibitem{newjapan} New Kyoto paper correcting NMR
%\bibitem{paulilimit}  Additional, older evidence of singlet pairing comes from the observation that the in-plane  critical  field at low $T$ appears to be Paul limited;  see ***
%
%\end{thebibliography}

\end{document}